\documentclass[5p,twocolumn,times,number]{elsarticle}

\usepackage{amsmath,amssymb}
\usepackage{bm}
\usepackage{graphicx}
\usepackage[dvipdfmx]{}
\usepackage{wrapfig}
\usepackage{hyperref}
\usepackage{lscape}
\usepackage{multirow}
\usepackage[caption=false,font=normalsize,labelfont=sf,textfont=sf]{subfig}
\usepackage{lineno,hyperref,subfiles,multicol}

\modulolinenumbers[5]
\journal{NIM-A}
\hypersetup{
colorlinks=true,
linkcolor=blue,
citecolor=blue,
urlcolor=blue,
}
\newcommand{\be}{\begin{equation}}
\newcommand{\ee}{\end{equation}}
\newcommand{\bea}{\begin{eqnarray}}
\newcommand{\eea}{\end{eqnarray}}

\newcommand{\tref}[1]{Table~\ref{#1}}
\newcommand{\eref}[1]{Eq.~(\ref{#1})}
\newcommand{\sref}[1]{Sect.~\ref{#1}}
\newcommand{\fref}[1]{Fig.~\ref{#1}}
\newcommand{\frefs}[1]{Figs.~\ref{#1}}

\begin{document}

\begin{frontmatter}

\title{The laser-based time calibration system for the MEG II pixelated Timing Counter}

\author[add1,add2]{G.~Boca}
\author[add1]{P.W.~Cattaneo}
\author[add3]{M.~De~Gerone}
\author[add4,add5]{M.~Francesconi}
\author[add4]{L.~Galli}
\author[add3,add6]{F.~Gatti}
\author[add7]{J.~Koga}
\author[add8]{M.~Nakao\corref{cor}}
\ead{nakao@icepp.s.u-tokyo.ac.jp, TEL: +81-3-3815-8384}
\author[add8]{M.~Nishimura}
\author[add9]{W.~Ootani}
\author[add1]{M.~Rossella}
\author[add9]{Y.~Uchiyama}
\author[add8]{M.~Usami}
\author[add8]{K.~Yanai}
\author[add8]{K.~Yoshida}

\cortext[cor]{Corresponding author}
\address[add1]{INFN Sezione di Pavia, Via A. Bassi 6, 27100 Pavia, Italy}
\address[add2]{Dipartimento di Fisica, Universit\'a degli Studi di Pavia, Via A. Bassi 6, 27100 Pavia, Italy}
\address[add3]{INFN Sezione di Genova, Via Dodecaneso 33, 16146 Genova, Italy}
\address[add4]{INFN Sezione di Pisa, Largo B. Pontecorvo 3, 56127 Pisa, Italy}
\address[add5]{Dipartimento di Fisica, Universit\'a degli Studi di Pisa, Via B. Pontecorvo 3, 56127 Pisa, Italy}
\address[add6]{Dipartimento di Fisica, Universit\'a degli Studi di Genova, Via Dodecaneso 33, 16146 Genova, Italy}
\address[add7]{Department of Physics, Kyushu University, 744 Motooka, Nishi-ku, Fukuoka 819-0395, Japan }
\address[add8]{Department of Physics, The University of Tokyo, 7-3-1 Hongo, Bunkyo-ku, Tokyo 113-0033, Japan}
\address[add9]{ICEPP, The University of Tokyo, 7-3-1 Hongo, Bunkyo-ku, Tokyo 113-0033, Japan}

\begin{abstract}
\small{
We have developed a new laser-based time calibration system for highly segmented scintillator counters like the MEG II pixelated Timing Counter (pTC), consisting of 512 centimetre-scale scintillator counters read out by silicon photomultipliers (SiPMs).
It is difficult to apply previous laser-based calibration methods for conventional metre-scale Time-Of-Flight detectors to the MEG II pTC from the implementation and the accuracy points of view.
This paper presents a new laser-based time calibration system which can overcome such difficulties.
A laser pulse is split into each scintillator counter via several optical components so that we can directly measure the time offset of each counter relative to the laser-emitted time.
We carefully tested all the components and procedures prior to the actual operation.
The laser system was installed into the pTC and thoroughly tested under the real experimental condition.
The system showed good stability and being sensitive to any change of timing larger than $\sim$10\,ps.
Moreover, it showed an uncertainty of 48\,ps in the determination of the time offsets, which meets our requirements.
The new method provides an example of the implementation of a precise timing alignment for the new type of detectors enabled by the advance of SiPM technology.
}
\end{abstract}

\end{frontmatter}


\section{Introduction}

Recently, timing measurements in experimental particle physics have improved in terms of two characteristics: flexibility of the detector design and the timing resolution.
Both aspects cannot be discussed separately and are related to the development of silicon photomultipliers (SiPMs).
The new improved timing detectors require dedicated methods for time calibration.
The challenges are the implementation in the complicated and multi-channel detectors and the accuracy of the calibration being sufficiently good compared to the improved detector intrinsic resolutions.

\begin{figure}[tb]
\centering
\includegraphics[width=\columnwidth]{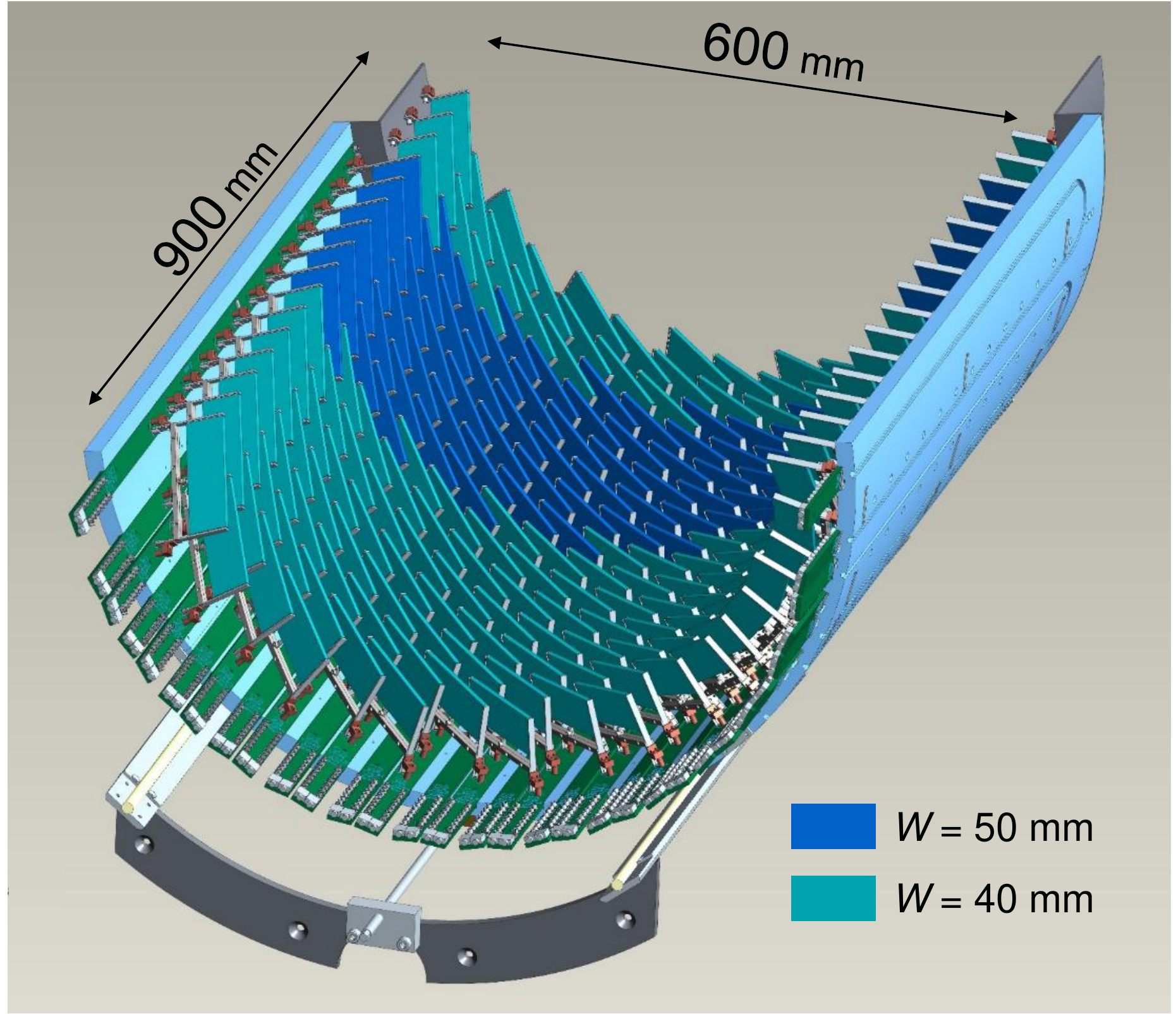}
\caption{Schematic view of the downstream pTC sector; the upstream sector is mirror symmetric (from \cite{Uchiyama2017}).}
\label{fig:pTC}
\end{figure}

The MEG II pixelated Timing Counter (pTC)\cite{Uchiyama2017} is such a new timing detector.
The MEG II experiment at Paul Scherrer Institut (PSI) in Switzerland will search for the lepton-flavour-violating muon decay, $\mu^+\to e^+\gamma$, with a branching fraction sensitivity of $6\times10^{-14}$\cite{MEG2}. 
The pTC is the subdetector dedicated to the measurement of the positron emission time.
It consists of two sectors (one placed upstream the target and the other downstream), each of which is composed of 256 scintillator counters read out by SiPMs as shown in \fref{fig:pTC}.
It achieves a time resolution of 38\,ps for the signal positrons by measuring the transit times with multiple counters.
In this paper, we present a new laser-based time calibration system developed for the MEG II pTC.

In the past decades, laser-based calibration methods were developed to calibrate and monitor timing detectors \cite{Brown1984, Kishida1987, Benlloch1990, Lacasse1998, Bonesini2003, Harris2008, Staric2017}.
However, it is difficult to apply similar methods to the MEG II pTC from the implementation and the accuracy points of view.
First, our detector is finely segmented in a complex geometrical configuration in limited space compared with conventional metre-scale Time-Of-Flight (TOF) detectors.
Scalability up to $\sim$500 channels and simplicity are required in the new approach.
We adopt the combination of an optical switch and optical splitters to divide the laser light instead of the conventional method using a diffuser.
This approach has higher scalability regardless of the original laser power and is more robust in distributing the light to each channel because a delicate optical system is not required.
We also developed a simple but robust method for the laser light injection to the counter, mechanically fitting the small counters.
Most of the optical components are commercially available and the system can easily be reproduced in other experiments.

Secondly, the precision and accuracy of the calibration are required to be good enough not to ruin the good detector resolution.
In the previous study\cite{Bertoni2016}, the dispersion of laser pulse in optical components is considered to be the main source of uncertainty.
In this paper, however, we discuss more on systematic uncertainties induced during the implementation of the system, which are found to be more important in actual operations.
We carefully tested all the components and procedures and measured the quantities necessary for the calibration, including the dispersion effect, prior to the actual operation of the system to suppress the uncertainty.

In the following sections, first, we overview the proposed calibration system and summarise each component in \sref{sec:system}. 
Then we report our R\&D work to complete the system design in \sref{sec:RandD}. 
In \sref{sec:pER2017}, the installation and operation of the system in the actual experimental environment are described and the performance is evaluated.
Finally we summarise our work in \sref{sec:conclusion}. 

\section{System overview}
\label{sec:system}
\subsection{MEG II pixelated Timing Counter}
\label{sec:pTC}
In the MEG I experiment, we achieved a positron time resolution of $\sim$70\,ps whereas its intrinsic time resolution was 40\,ps\cite{MEG2}.
This degradation was because of a large variation of the optical photon paths due to the large scintillator size ($80\times4\times4\,$cm$^3$), the worse PMT performance under the magnetic environment, and an incomplete time calibration.
To overcome such limitations, we upgraded the timing counter for the MEG II experiment.
We adopted a highly segmented design with 512 scintillator counters divided into two mirror symmetric sectors.
The design was determined to maximise the experimental sensitivity under a limited number of electronics readout channels by a Monte-Carlo study.
Multiple measurements with the smaller scintillator counters enable us to reach $\sim$38\,ps time resolution.
In addition, this multiple hit scheme makes the timing measurement less sensitive to the error of timing alignment and the electronics time jitter.

\begin{figure}[tb]
\centering
\includegraphics[width=\columnwidth]{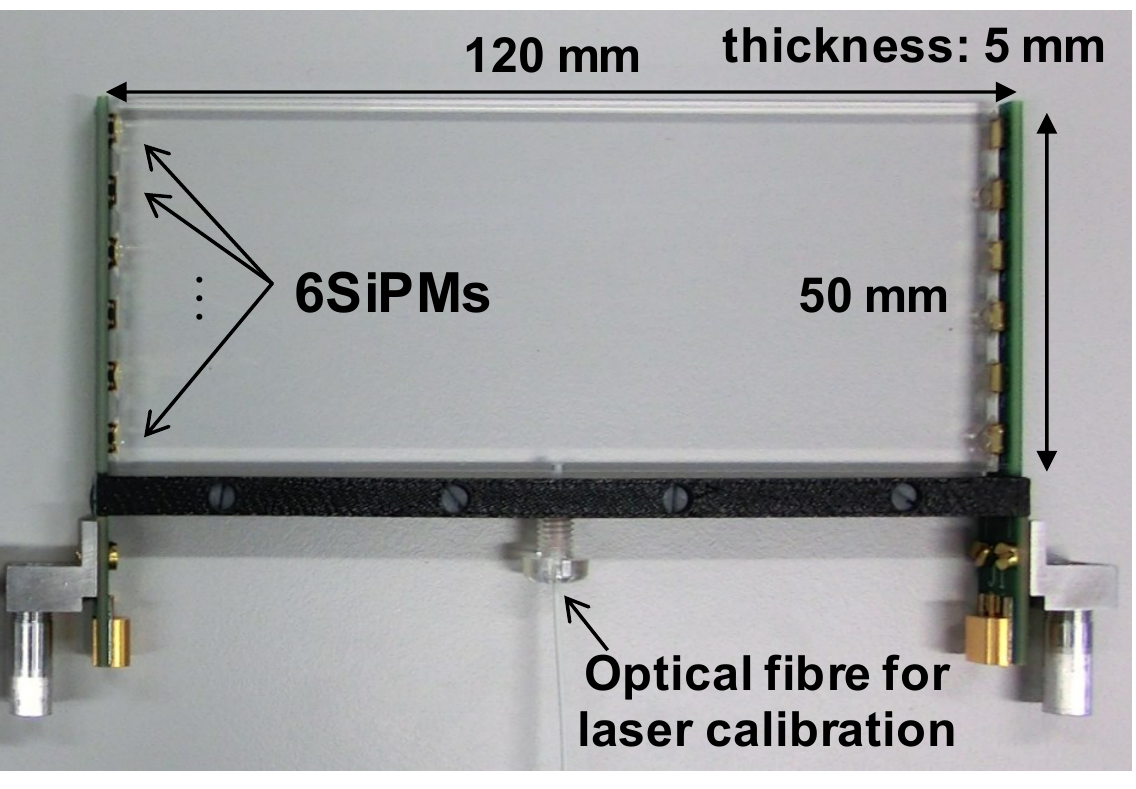}
\caption{A single scintillator counter with height 50 mm. The scintillator will be wrapped in a reflector and then wrapped in a black sheet for light shielding (see the text for details).}
\label{counter}
\end{figure}

\begin{figure}[tb]
\centering
\includegraphics[width=\columnwidth]{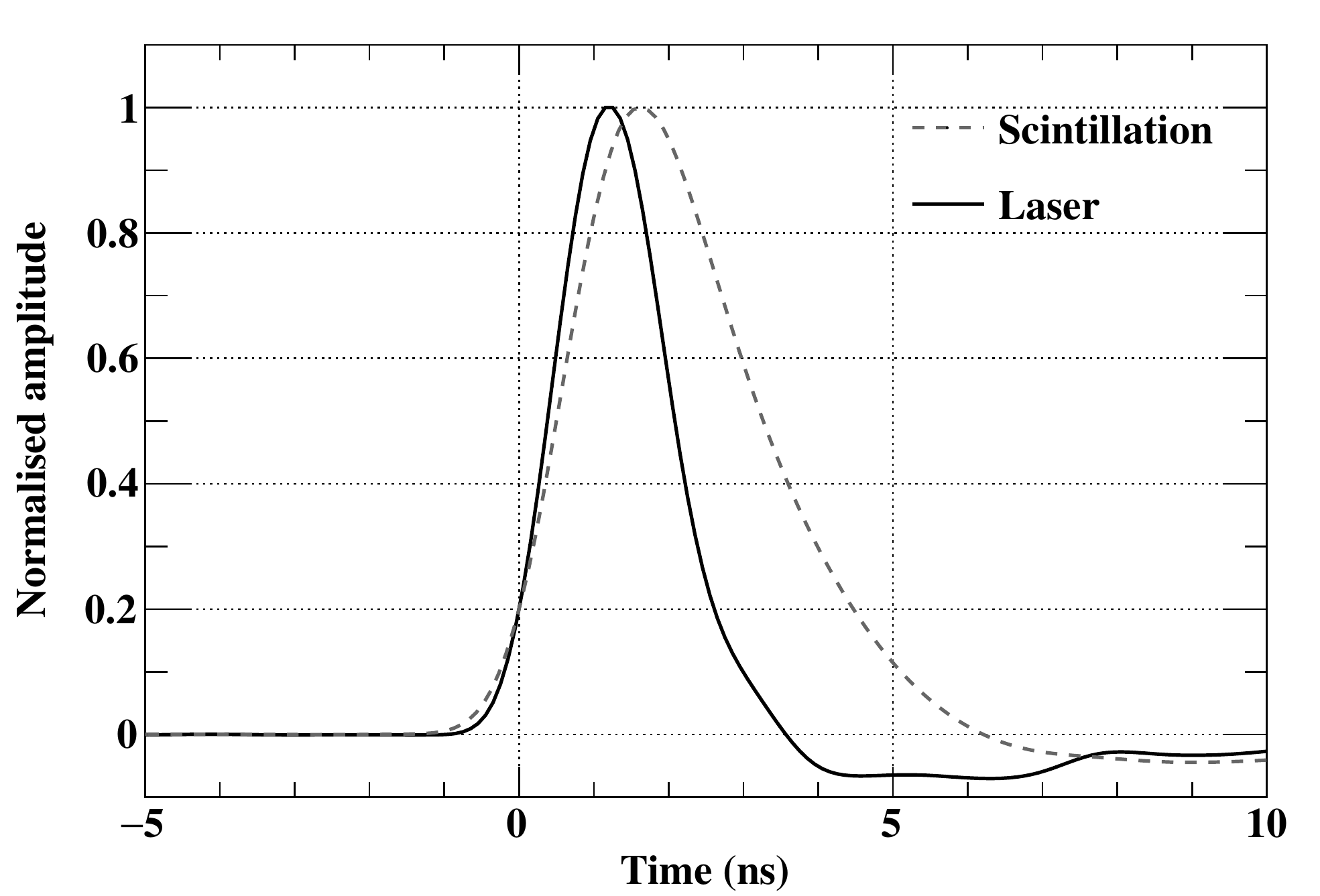}
\caption{Averaged waveforms of the laser signal (solid line) and the scintillation signal (broken line) read out by the six series-connected SiPMs after shaping with pole-zero cancellation. 
The amplitudes are scaled to 1.
The constant fraction of 20\% is used for timing calculation and thus $(\mathrm{Time}, \mathrm{Normalised\,amplitude}) = (0, 0.2)$ is fixed for comparison.
}
\label{waveform}
\end{figure}

The single scintillator counter is made of a fast plastic scintillator, BC422 with dimensions of $40\, \mathrm{or}\,50\times120\times5\,\mathrm{mm}^3$ read out by six series-connected SiPMs at both ends as shown in \fref{counter}.
The peak wavelength of the scintillation emission is 370\,nm\cite{SaintGobain}.
The rise time and decay time of the scintillator are 0.35\,ns\footnote{This value is dominated by the measurement setup and the intrinsic one is much faster\cite{Lerche}.} and 1.6\,ns, respectively\cite{SaintGobain}.
The scintillator is wrapped in 32\,$\mu$m thick ESR2 film, which has a reflectivity higher than 98\%\cite{ESR} across the visible spectrum (400--800 nm).
The SiPMs are produced by AdvanSiD and most of them (92\%) are ASD-NUV3S-P High-Gain (MEG) and the others are ASD-NUV3S-P\footnote{These two models were made in different production lots. The latter one, ordered about a year later when the production lot of the former was already closed, has a wider voltage range for its operation.}.
The active area and pixel pitch are 3 $\times$ 3 mm$^2$ and 50 $\times$ 50 $\muup\mathrm{m}^2$, respectively.
The peak wavelength of the photon detection efficiency (PDE) is 420\,nm.
The breakdown voltage is $\sim$24\,V and the operational voltage for a six series-connected chain is $\sim$164\,V. 
The optical fibre for the laser-based time calibration is inserted from the bottom.
The counters are individually wrapped in a 25\,$\mu$m-thick black sheet (Tedlar$^{\textregistered}$\cite{DuPont}) for light shielding.
At each side, $\sim$50 photoelectrons are detected for a traverse of a minimum ionising particle such as a 50\,MeV positron.
The typical waveforms for the scintillation signal and the laser signal are shown in \fref{waveform}.
The laser signal has a sharper shape than the scintillation signal:
the rise times of laser and scintillation are measured to be $1.10\pm0.04$\,ns and $1.41\pm0.01$\,ns, respectively, where the rise time is defined as time between 10\% and 90\% of its height.
The FWHMs are $1.84\pm0.12$\,ns and $2.75\pm0.65$\,ns, respectively\footnote{Setup(3) in \sref{sec:setup} was used for this measurement. We averaged over $\sim$300 counters to calculate these values.}.
These differences should be taken into account in calculating the time offsets in \sref{sec:pER2017}.

The positron hit time at each counter is calculated as follows:
\bea
t_{\mathrm{hit}}&=&\frac{t_1+t_2}{2}-\frac{l}{2v_{\mathrm{eff}}}-t_{\mathrm{elec}},
\label{eq:hit}
\eea
where $t_1$ and $t_2$ are the measured signal time on channel 1 and channel 2, respectively; $l = 120$\,mm is the longitudinal length of the counter; $v_{\mathrm{eff}}\simeq12.4$\,cm/ns is an effective velocity of the light in the scintillator; and $t_{\mathrm{elec}}$ is a signal transmission time in electrical components such as cables and DAQ line.
The sum of the last two components is defined as the time offset ($t_{\mathrm{offset}}$) of the counter to be calibrated:
\bea
t_{\mathrm{offset}} = \frac{l}{2v_{\mathrm{eff}}}+t_{\mathrm{elec}}.
\label{timeoffset}
\eea
The relative value of the time offset among the scintillator counters matters in the time calibration proposed in this paper.

\subsection{Time calibration}
To determine the time offset in \eref{timeoffset}, we have developed two complementary time calibration methods: the track-based method and the laser-based method.
Positron tracks from the Michel decays ($\mu^+\to e^+\nu_e\bar{\nu}_{\mu}$) are used in the track-based one and laser signals are used in the laser-based one.
These methods are combined to determine the time offsets.
The comparison between the two methods is summarised in \tref{TimeCalib}.
The laser-based method has advantages over the track-based method in terms of position dependence, DAQ time, and the beam requirement.
On the other hand, it does not have 100\% coverage because counters placed on the very inner part of the detector do not have enough space to insert any other items including laser fibres.
In the development of the laser-based method, it is important to minimise its uncertainty retaining the advantages over the other method.

\begin{table}[tb]
  \centering
  \small
   \caption{Comparison between two time calibration methods}
   \label{TimeCalib}
    \begin{tabular}{llc} \hline
      \textbf{Item} & \textbf{Laser} & \textbf{Track}  \\ \hline \hline
      Position dependence & \textbf{No} &Yes \\ 
      DAQ time & \textbf{short} ($\sim30$\,min.)  &long ($\sim$2 days) \\ 
      Beam & \textbf{not necessary} & necessary \\
      Coverage & 84\% & \textbf{100\%} \\ 
      Uncertainty & 27\,ps & \textbf{13\,ps (simulation)}  \\ 
      \hline
    \end{tabular}
\end{table}

\subsection{The laser calibration system}
\subsubsection{Optical components}
\label{sec:laser_system}

\begin{figure*}[tb]
\centering
\subfloat[]{\includegraphics[width=\columnwidth]{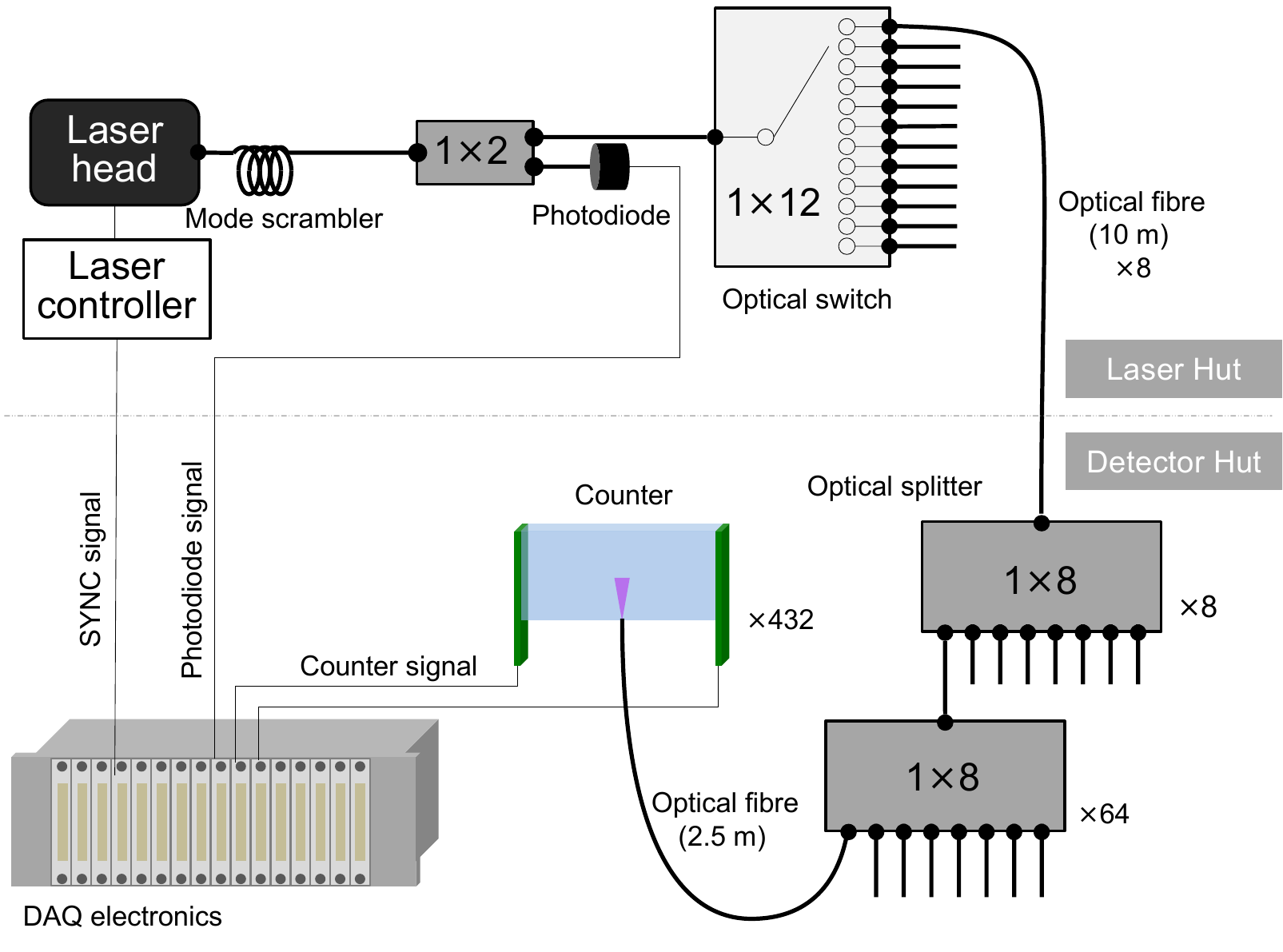}
\label{laser_system_a}}
\hfil
\subfloat[]{\includegraphics[width=\columnwidth]{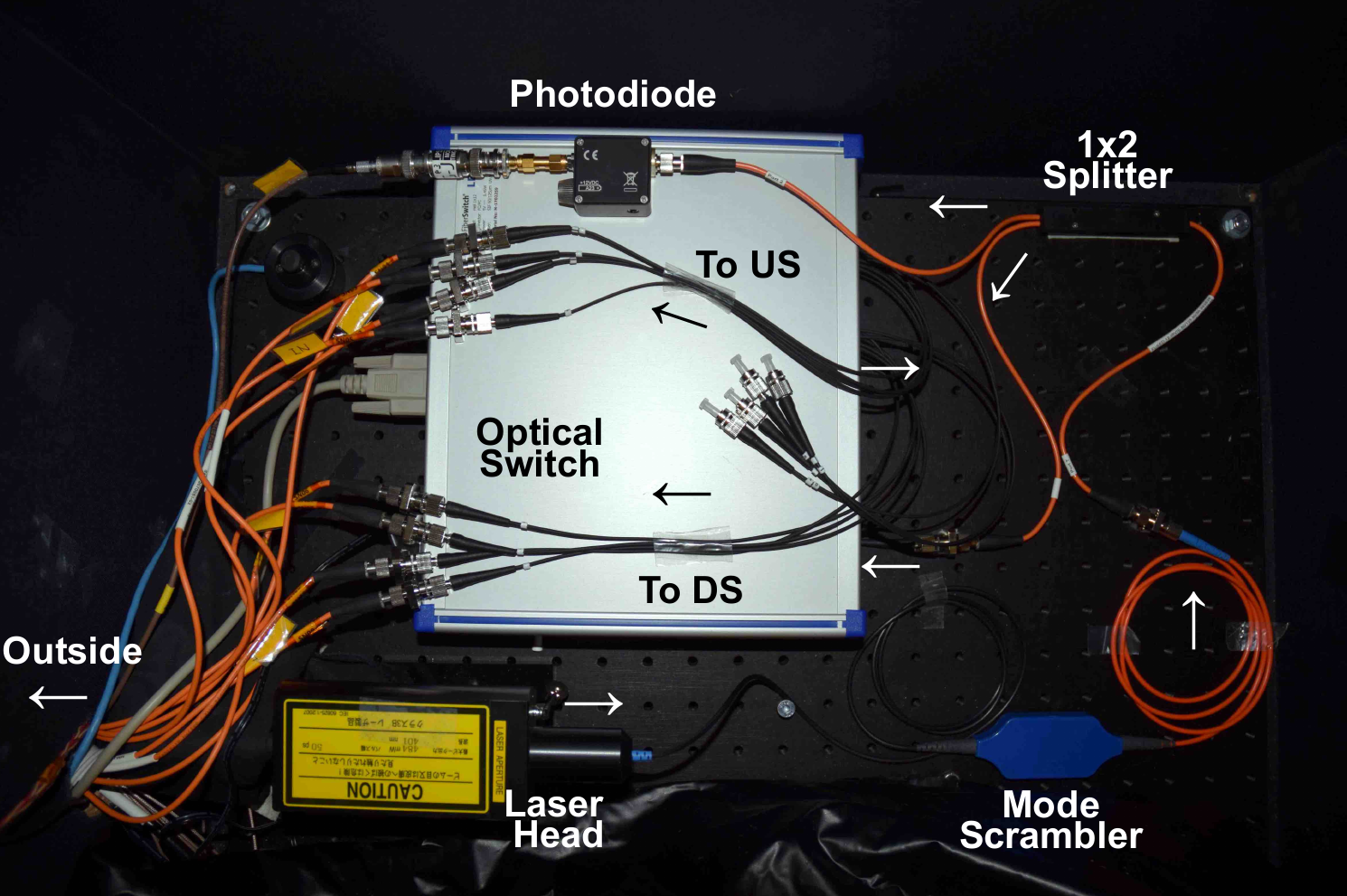}
\label{laser_system_b}}
\caption{(a)Schematic and (b)picture (from \cite{Nakao2019}) of the laser-based time calibration system for the MEG II pTC. 
The individual components are described in the text.
}
\label{laser_system}
\end{figure*}

The laser calibration system is shown in \fref{laser_system}.  The laser light is split along a ladder of optical components into $>\!400$ channels and distributed to the counters via optical fibres.
The optical components were selected based on the detailed study presented in \cite{Bertoni2016} and are summarised in \tref{laser_list}.

\begin{table*}[t]
  \centering
  \small
   \caption{Optical components used in the laser calibration system}
   \label{laser_list}
    \begin{tabular}{lllll} \hline
	\textbf{Item} & \textbf{Model} & \textbf{Specifications}  & \textbf{Pieces} & \textbf{Ref.} \\
	                      &  (Manufacturer) &                        & (spares)  &
       \\ \hline \hline
      laser & Picosecond Light Pulser PLP-10 & Wavelength 405~nm,  pulse width 60  ps,  & 1 & \cite{laser} \\
      & (HAMAMATSU PHOTONICS K.K.) &    peak power 200 mW (typical values)     & & \\

      optical fibre (2.5\,m) & QMMJ-31-IRVIS-50/125-1HYWT-2.5-SP &  High power GI multimode, core/clad 50/125 $\muup$m,  & 432 & \cite{fibre25} \\
      & (OZ OPTICS) &NA 0.2, 400--2000 nm, 1 mm jacket  &  (+129) & \\
      
      optical fibre (10\,m) & MMJ-33-IRVIS-50/125-3-10 & GI multimode, core/clad 50/125 $\muup$m, & 8 & \cite{fibre10} \\
      & (OZ OPTICS) &NA 0.2, 400--2000 nm & (+6) & \\ 
       
      mode scrambler & ModCon Mode Controller & Insertion loss $< 3$ dB at 850 nm & 1 & \cite{mode} \\
      & (Arden PHOTONICS) & & & \\
%
      
      1$\times$2 splitter & FUSED-12-IRVIS-50/125-50/50-3S3S3S-3-0.25 & FBT multi-mode coupler,  400--1600 nm,  & 1 & \cite{2splitter} \\
      & (OZ OPTICS) & excess loss $< 1.0$ dB for 480--700 nm& & \\
      
      1$\times$8 splitter & MMC-18-A-EVEN-1-A-30CM-R-1 &  FBT multi-mode coupler, insertion loss $\le 11.5$ dB & 70 & \cite{8splitter} \\
      & (Lightel Technologies Inc.) & & (+11) &  \\
      
      photodiode & DET02AFC & 400--1100 nm, bandwidh 1 GHz, rise time 1 ns& 1 & \cite{PD} \\
      & (THORLABS) & & & \\
      
      optical switch & fibre Optical Switch mol 1$\times$12-50\,$\muup$m & Insertion loss $<$  2.0 dB for 5--16 output channels& 1 & \cite{switch} \\
      &  (LEONI)  & & & \\      
%
      \hline
    \end{tabular}
\end{table*}

We use a picosecond pulsed diode laser as the light source. The device emits a short light pulse, a duration of 50\,ps (FWHM), at a wavelength of 401\,nm. 
The maximum peak power is 484\,mW, equivalent to $\sim\!10^7$ photons per pulse (measured).
The ``laser controller" controls the ON/OFF switching, the repetition rate (up to 100~MHz), and the power.
We operate it at a power $\sim$1/3 of its maximum.
The laser power is monitored by the photodiode connected to one of the outputs of the first stage 1$\times$2 splitter.

The laser controller also provides a pulse signal\footnote{either NIM or TTL signal} synchronised with the laser emission, with an adjustable delay.
We call this signal the ``SYNC signal'' and use it for the trigger and the timing reference.
The jitter between the SYNC signal and the optical output is less than 10~ps according to the specification and is negligibly small for the usages.
By using this reliable time reference, it is possible to divide the counters to be illuminated at a time into several subsets.

We choose multi-mode fibres with a graded index (GI) for the light transmission because
they have less insertion loss than single mode fibres thanks to their larger core diameter (50~$\muup$m).
However, optical paths in multi-mode fibres differ depending on the optical modes.
The graded refractive index mitigates the effect on the dispersion of propagation time.
The mode scrambler is used to stabilise the optical mode propagation in the multi-mode fibres by removing high-order modes within a wound fibre.

The optical switch actively switches from one output channel to another using a micro-mechanical technology. It has 12 output channels and 8 of them are used for the calibration while the others are kept as spares.
The components up to the switch (shown in \fref{laser_system_b}) are placed in a laser hut next to the detector hut.

The optical splitters passively split the input signal into the output channels using the Fuse Biconical Taper (FBT) technology~\cite{Kawasaki1977}.
Two cascaded stages of 1$\times$8 splitters are used to simultaneously deliver the laser pulse to up to 64 counters.
The combinations were selected to optimise the output uniformity.
About 40 splitters connected to each (upstream or downstream) sector of the pTC are housed in a box placed close to the detector.  

The final component to inject the laser pulse into the counter is the 2.5\,m fibre.
The far end of the fibre is covered with a ceramic (zirconia) ferrule, without any magnetic metal sleeve, to allow operation in a high magnetic environment.

FC/PC connectors are used at all the connections between the components.

In total, the light power is attenuated to $\sim\!10^{-4}$ of the original, resulting in thousands of photons are delivered to each counter; and finally, $\sim\!50$ photoelectrons are detected by the six SiPMs at each end.

\subsubsection{Monitoring and controlling}
Slow control tasks such as monitoring temperatures and controlling the laser, and the optical switch are performed remotely based on the Maximum Integrated Data Acquisition System (MIDAS)\footnote{MIDAS provides several tools and a general program framework to assist experiments. An example of the usage can be found in \cite{Adam2013}.}~\cite{MIDASwebsite} developed at PSI and TRIUMF.
Temperature sensors, the laser controller, and the optical switch are monitored and controlled by the system remotely.

\subsection{Readout electronics and data acquisition}
\label{sec:daq}
The counters are mounted on 1-m long PCBs, which have a 50 $\Omega$ characteristic impedance.
The signals are transmitted on RG--178 coaxial cables ($\sim$35\,ns), which connect to custom DAQ boards.
A multi-functional DAQ board called WaveDREAM2~\cite{wavedream, wavedream2} has been developed at PSI for the MEG II detectors, including pTC.
It has the following functionalities:
SiPM biasing, amplification, shaping with pole-zero cancellation, waveform digitisation at GSPS using Domino Ring Sampler (DRS4) chip\cite{DRS}, and the first level trigger.
The SYNC and the photodiode signals, as well as the counter signals, are fed into the boards and digitised. 
The SYNC signal is used to trigger the laser data.

\section{Research and development}
\label{sec:RandD}
\subsection{Experimental setup}
\label{sec:setup}
\begin{figure}[!t]
\centering
\includegraphics[width=\columnwidth]{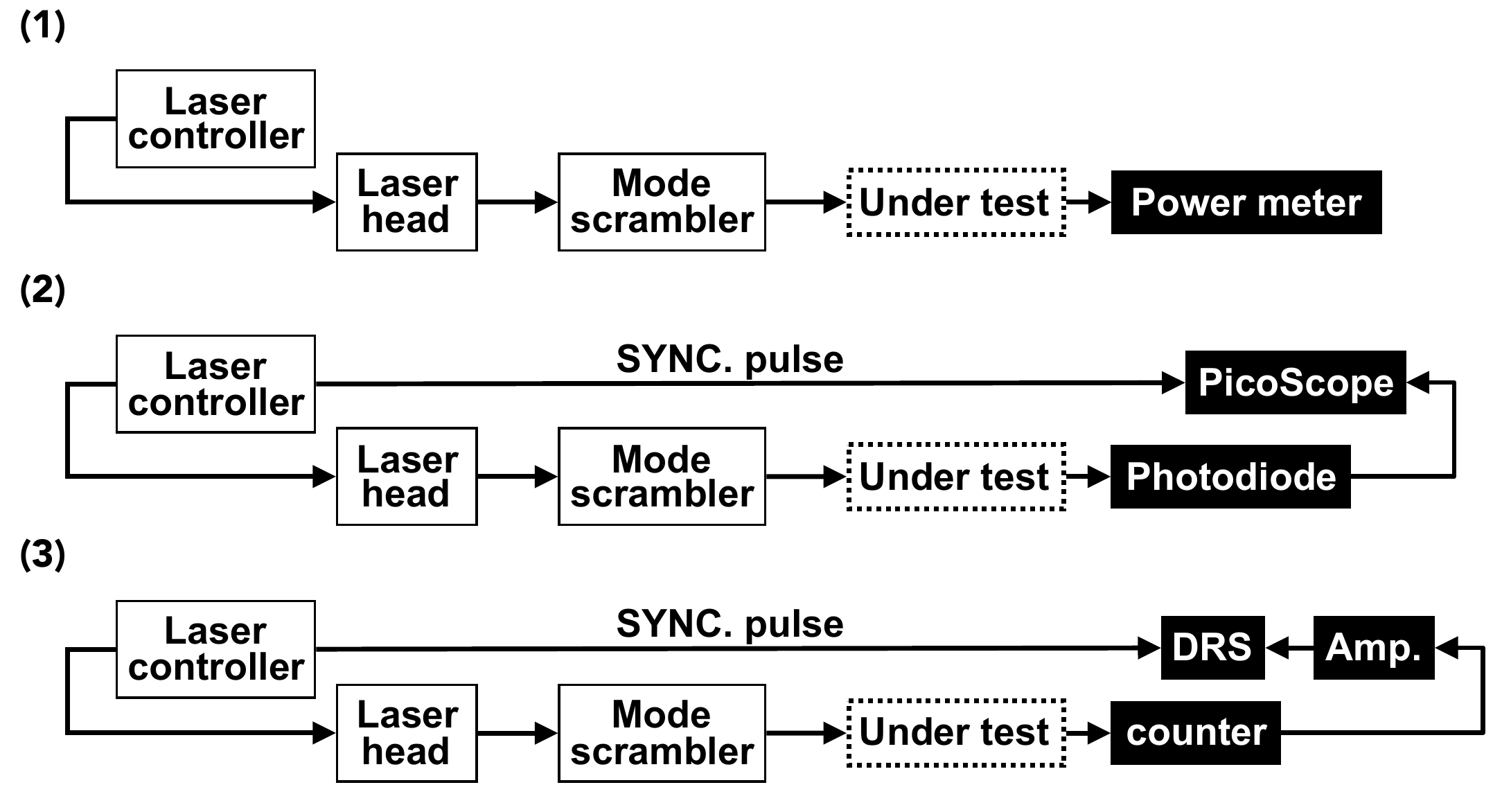}
\caption{Experimental setups used in the research and development phase.}
\label{testsetup}
\end{figure}

The setups used in the laboratory measurements are shown in \fref{testsetup}.
We used three different setups depending on the purpose:
\begin{enumerate}
\setlength{\leftskip}{1.0cm}
\item [Setup (1)]
 To measure the power of laser light at each point and the insertion loss of the component under test, a power meter\footnote{Model 840-C HandHeld Optical Power Meter by Newport} is used.
\item [Setup (2)]
To measure the optical length of the component under test, a PC-connected oscilloscope\footnote{PicoScope9210 (bandwidth: 12 GHz, sampling: 100 GS/s) by Pico Technology} is used. The lengths are measured as the delay time with respect to the SYNC signal.
\item [Setup (3)]
To acquire the waveform signal from the scintillator counters, a waveform digitiser\footnote{DRS Evaluation board V.~4 developed at Paul Scherrer Institut, sampling speed was set to 1.6 GS/s} and a SiPM amplifier are used\footnote{Bandwidth of the full chain: 440 MHz}. The SYNC signal is used for the trigger and the time reference.
\end{enumerate}


\subsection{Light injection and fibre fixing method}
\label{fibre_insertion}
In the past, several approaches were used to couple an end fibre to a scintillator:
simply attaching the fibre to the scintillator face beside a photo sensor~\cite{Kishida1987, Harris2008} or using a prism~\cite{Brown1984, Benlloch1990, Lacasse1998, Bonesini2003, Staric2017}. 
We have developed a novel approach to fix a fibre in order to reduce materials and cost and to achieve good properties described in this section.

\begin{figure}[!t]
\centering
\includegraphics[width=\columnwidth]{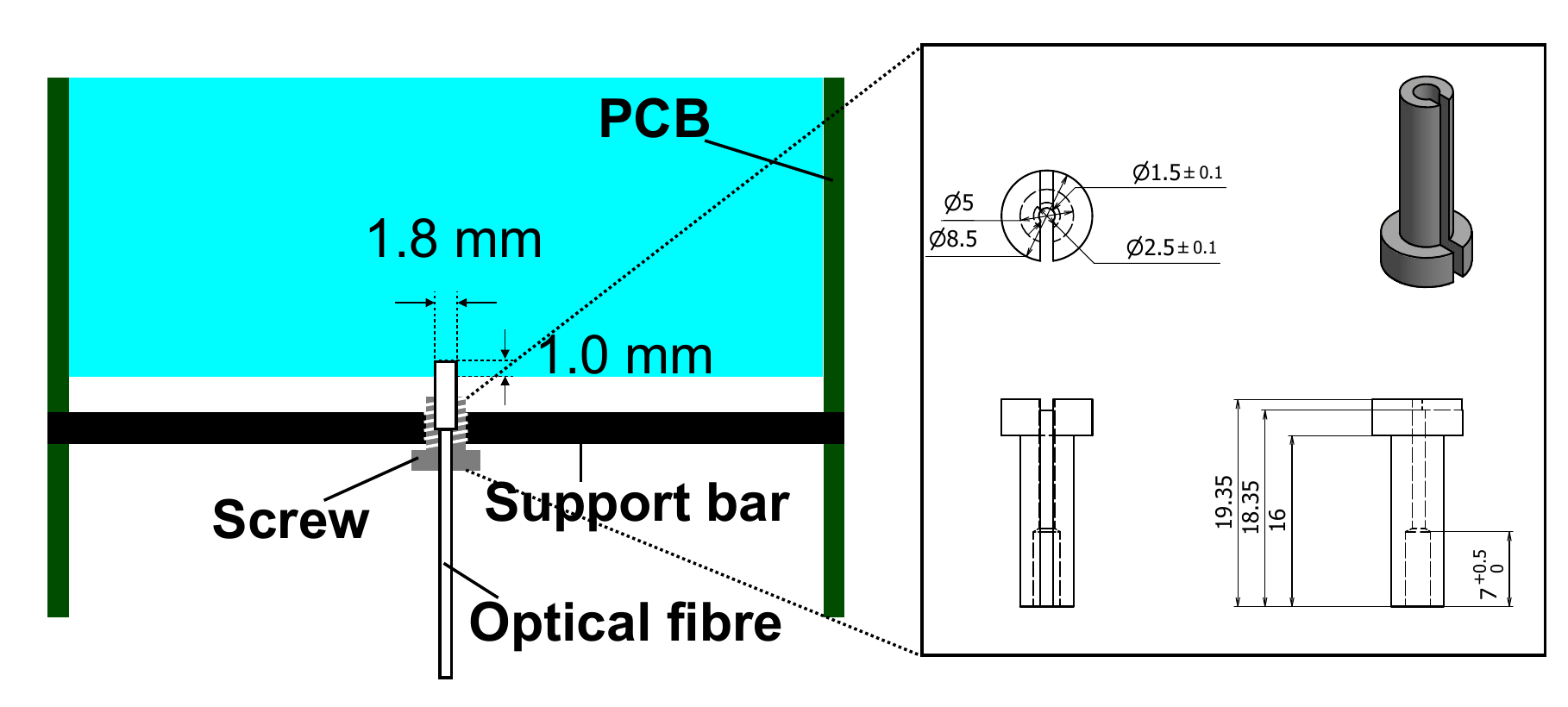}
\caption{A cross-sectional view of fibre fixing method (left) and the design of the screw. The screw has a slit to insert the fibre and dedicated hole to fix the fibre (right).}
\label{fibre_fix_method}
\end{figure}

\fref{fibre_fix_method} illustrates our method for inserting a fibre into the scintillator counter.
The laser light is vertically incident on the centre of the bottom face of the scintillator.
The fibre tip is fixed in two ways: the top is fixed by a hole on the scintillator, and the bottom is fixed by a specially machined screw (polycarbonate) supported by a bar (ABS resin) across the two PCBs.

Since this is the most delicate part of the system, we performed several measurements to examine the functionality of this method. 

\paragraph{Effect of the hole on the scintillator}
The size of the hole was minimised (2.5~mm diameter, 1~mm depth) to avoid any negative effect on the counter performance.
To study the effect on the scintillation light collection efficiency, we first made a Geant4-based Monte-Carlo simulation.  
The result did not show a significant decrease in the light collection efficiency and thus in the time resolution.
Then, we measured the time resolution of a few test counters using electrons from a $^{90}$Sr source impinging at the centre of the counter before and after drilling the holes and confirmed that the hole does not affect the time resolution\footnote{see \cite{Cattaneo2014} for the set up of the time resolution measurement.}.

\paragraph{Coupling method}
We tested the coupling between the fibre and the scintillator under the following two cases: in air and with optical grease.
The time centre, defined as the mean of measured distribution of $(t_1+t_2)/2-t_{\mathrm{SYNC}}$, turned out to be dependent on the amount of the grease put in the hole.
The amount of the laser light observed at the photosensor can be changed by 4 times at the maximum and the time centre changed by 300\,ps depending on the amount of the grease although we tried to put the same amount at every measurement.
Since it is difficult to control the amount of grease with sufficient precision during the assembly,
we decided not to use optical grease. 

\paragraph{Asymmetry between two channels}
We also checked the splitting ratio of laser light between channel 1 and channel 2.
The ratio $R_{\mathrm{ch1}}$ defined as 
\bea
R_{\mathrm{ch1}}= \frac{A_1}{\epsilon_1}\bigg/ \left(\frac{A_1}{\epsilon_1}+\frac{A_2}{\epsilon_2}\right)
\eea
is measured to be $(50\pm4)\%$, where $A_i$ denotes the signal amplitude and $\epsilon_i$ is the product of the gain and the PDE of the SiPMs.
PDE was relatively calculated 
from the scintillation signal data obtained by irradiating the centre of counters with electrons from a  $^{90}$Sr source.
The injection method showed good splitting ratio and needs no correction.

\paragraph{Reproducibility}
\label{sec:reproducibility}
One of the important points is the reproducibility of the light injection; the time centre should be stable after reinserting the fibre. 
To test it, we repeated the timing measurement 10 times by extracting and inserting the fibre every time.
The standard deviations of the time centre for 4\,cm counter and 5\,cm counter were 11\,ps and 3.0\,ps, respectively.
The lower variation for 5\,cm counter stems from the fact that the distance between fibre support and the hole on the scintillator is shorter than that of 4\,cm by 1\,cm and thus the tip of the fibre is fixed more accurately. 
Note that these measured values are well below our requirement.

\paragraph{Stability}
The stability was first checked using a test bench for 70 hours and the standard deviation of the time centre was 9.2\,ps, which is sufficient for our use.
Then it was confirmed in the commissioning phase after installation of the system (Sect.~\ref{sec:pER2017}). 

\subsection{Timing accuracy, precision, and dispersion}
The timing property is determined by the number of detected photoelectrons and the arrival time distribution.
As shown in \fref{waveform}, the arrival time distribution of laser light is narrower than that of scintillation light, resulting in a good timing accuracy.

Nevertheless, a finite width of the distribution would cause a systematic error on the detected timing of the laser signal.
In \cite{Bertoni2016}, the dispersion of laser pulse in optical components is considered to be the main source of the error.
To evaluate the impact on the output signal, we compare the signals when we inject the laser pulse via only an adjustable optical attenuator and when we do via the attenuator and the full sequence of the optical components.
The rise times are measured\footnote{An oscilloscope, DPO 4104 (bandwidth: 1 GHz, sampling: 5 GS/s) produced by Tektronix was used.} to be $975\pm66$\,ps and $965\pm77$\,ps, respectively; no significant difference was observed, validating our choice of the optical components with a multi-mode network.

An optical simulation indicates that the photon timing distribution is predominantly determined by the dispersion inside the scintillator.
The fast rise time is given by direct or a few times reflected photons while the width is given from photons reflected many times.
The counter-by-counter variation of the distribution causes a systematic error if we do not know the impact.
Therefore, we measure the effective optical length of the full sequence of the optical components, including the scintillator, for all the counters in the mass test described in \sref{mass_test}.
The effective optical length includes the effect of the dispersion and thus the systematic error is eliminated by subtracting the effective optical length from the time centre.

The time resolution for the laser signal is 50\,ps (standard deviation of a fitted Gaussian), better than that for the scintillation signal.
The time offset is determined from the mean of the Gaussian and thus the precision can be improved with the statistics.
For example, 1\,ps precision is achievable with 3000 events.

\subsection{Temperature dependence}
A temperature variation may change the optical and electrical length of the system components.
Therefore, we measured their temperature coefficients.
The results are summarised in \tref{TemperatureDependence}.
In these measurements, only the item under test was put in a thermal chamber\footnote{Bench-Top Type Temperature and Humidity Chamber SU-241 by ESPEC} and the others were kept in the room temperature.

Originally, we used an RG174/U cable for the SYNC signal transmission, but it turned out to be the dominant source of the temperature dependence of the system.
Therefore, we replaced it with a less-temperature-sensitive cable, FSJ1-50A (COMMSCOPE)\cite{FSJ1}. 
It is designed to have a linear expansion of the conductor part canceling out the temperature dependence of the dielectric constant of the insulator part.
As a result, it has a small temperature coefficient of $-0.08$ ps/K, two orders of magnitude smaller than that of RG 174/U ($-8.8$ ps/K).

We measured the temperature coefficient of the whole system to be $-1.3\pm0.2$\,ps/K.
The temperature in the detector hut is expected to be stabilised within 1 K by an air conditioner.
Thus, the temperature effect on the time centre is negligible.

\begin{table}[tb]
  \centering
  \small
   \caption{Summary of temperature dependence.
   The last column reports the sign of the effect on the time offset.}
   \label{TemperatureDependence}
    \begin{tabular}{llc} \hline
      \textbf{Item} & \textbf{Coefficients\,(ps/K)} & \textbf{Effect}  \\ \hline \hline
      Fibre (2.5\,m) & $+0.45\pm0.02$ &+ \\ 
      Fibre (10\,m) & $+1.00\pm0.04$  &+ \\ 
      Optical splitter & $+0.24\pm0.11$ & + \\ 
      Scintillator counter & $+1.24\pm0.04$ & + \\ 
      Counter signal cable & $-4.1\pm0.1$ & $-$ \\ 
      SYNC signal cable & $-0.08\pm0.02$& +\\ \hline 
      Total & $-1.3\pm0.2$ & $-$ \\ 
      \hline
    \end{tabular}
\end{table}

\subsection{Mass test}
\label{mass_test}
We thoroughly tested all the optical components in terms of optical length and the ratio of output to input power ($R_\mathrm{out}$)\footnote{The insertion loss is given by $-10\log_{10}R_\mathrm{out}$.}.
Since the optical lengths of all the optical components are not controlled with the required precision,
we need to measure them beforehand as a pre-calibration and then subtract them when we determine the time offsets.
The overall lengths measured with the final combinations of the components are used for this purpose, while the lengths of individual components can be used when we replace some parts in the future.
The results of $R_\mathrm{out}$ measurement are used in making the combination of these items to make the outputs uniform.

\paragraph{$\mathit{1\times2}$ splitter}
The difference of optical length between channel 1 and channel 2 is measured to be $17.0\pm0.6$\,ps (channel 2 is longer) while the absolute optical length is $\sim\!3.1$\,ns.
The $R_\mathrm{out}$'s are measured to be $26.6\pm0.3$\% (channel 1) and $31.6\pm0.3$\% (channel 2).

\begin{figure}[!t]
\centering
\includegraphics[width=\columnwidth]{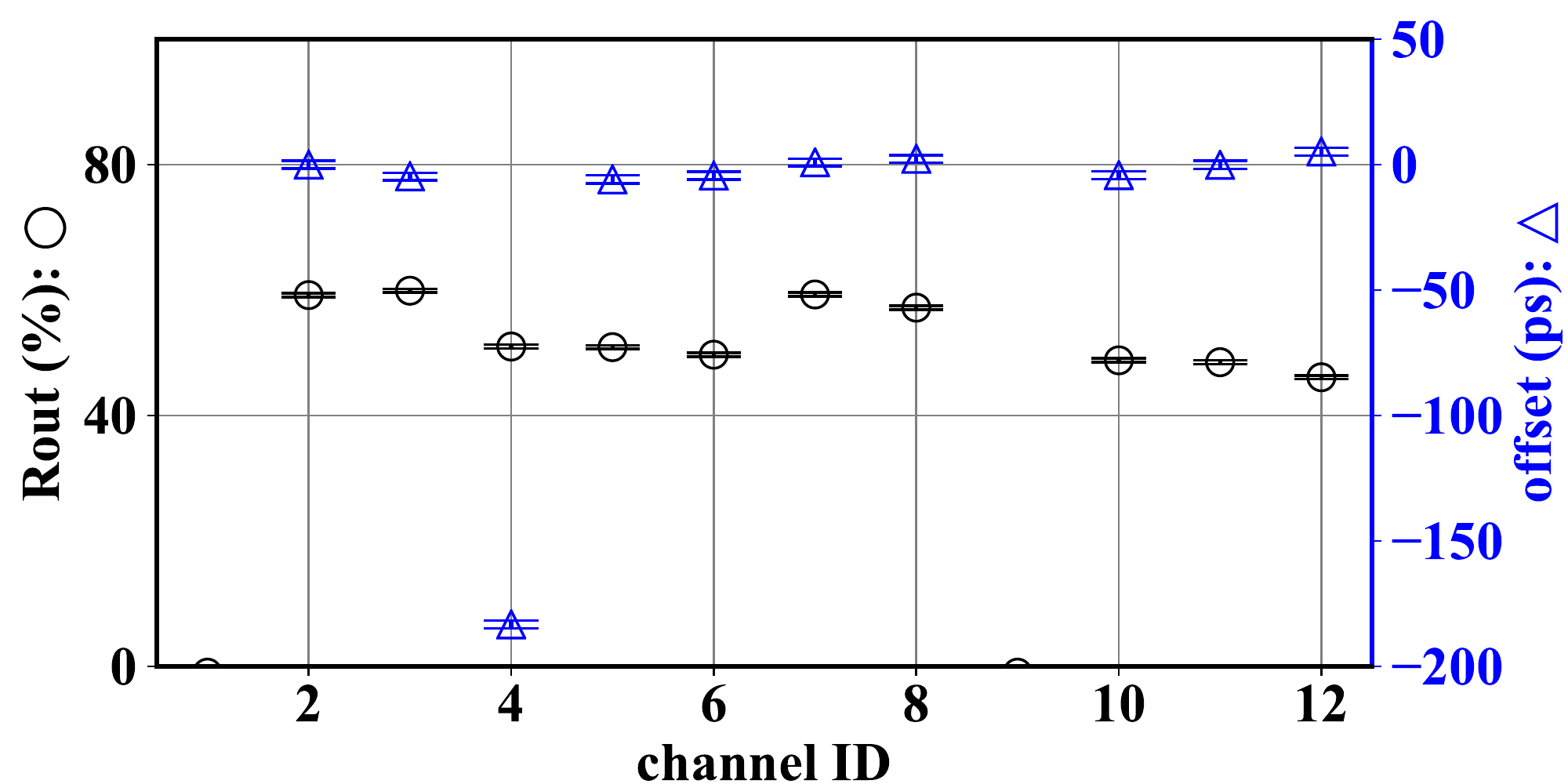}
\caption{$R_\mathrm{out}$ (circle) and optical length difference (triangle) of the optical switch. 
The optical length differences are given with respect to channel 2. Channel 1 and 9 are broken.}
\label{optical_switch_py}
\end{figure}

\paragraph{Optical switch}
\fref{optical_switch_py} shows the results for the optical switch.
The $R_\mathrm{out}$'s are measured to be 50--60\%.
The differences of the optical lengths are less than 6\,ps but for channel 4.
This channel has a larger difference ($\sim\!200$\,ps) because it was originally broken after its shipment and repaired by the manufacturer.
Channel 1 and 9 have no data because they were broken.

\begin{figure}[!t]
\centering
\includegraphics[width=\columnwidth]{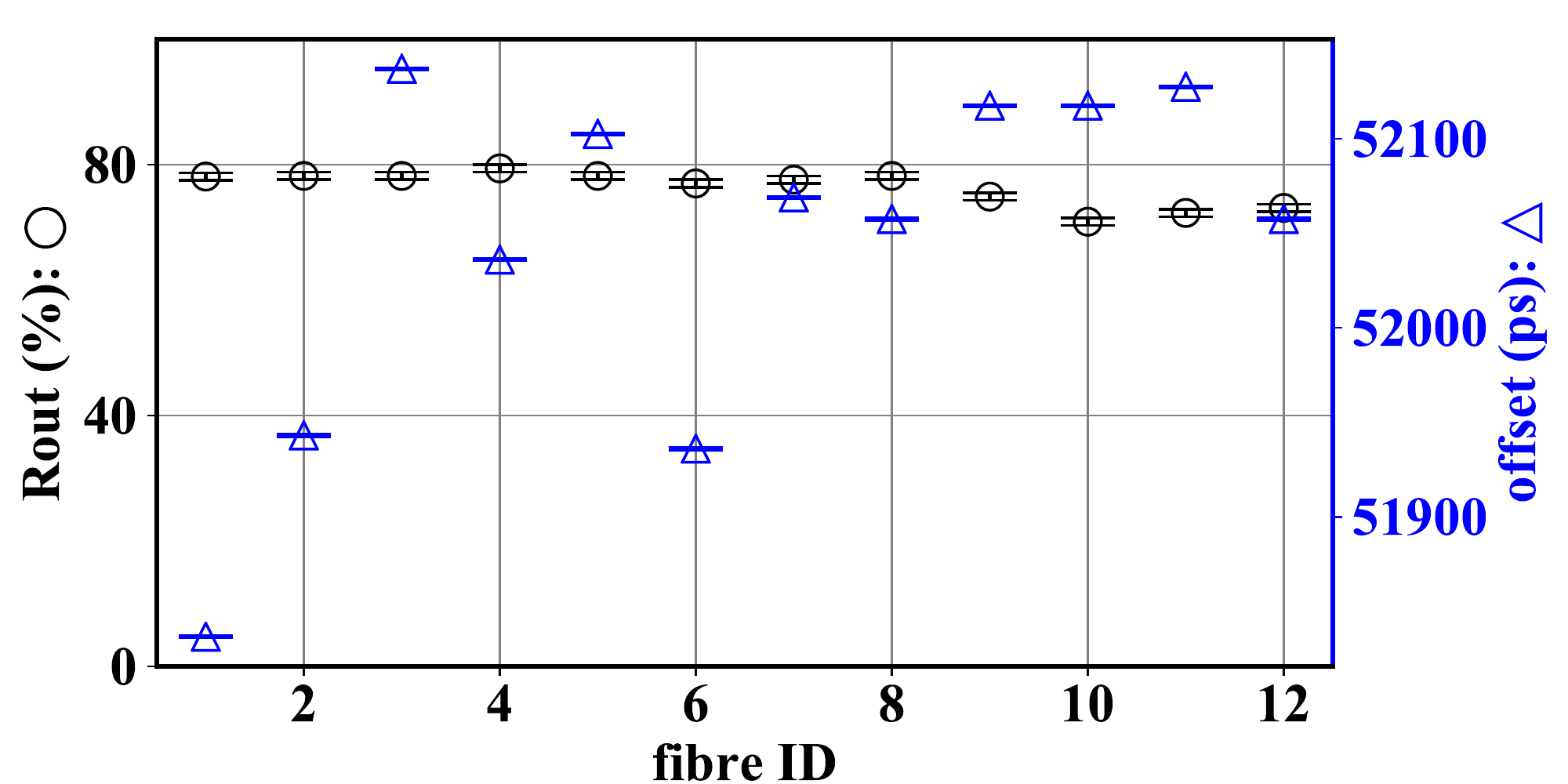}
\caption{$R_\mathrm{out}$ (circle) and optical length difference (triangle) of the long fibres.}
\label{long_fibre_py}
\end{figure}

\paragraph{Long fibres}
\fref{long_fibre_py} shows the results for the long fibres.
The $R_\mathrm{out}$'s are measured to be 70--80\%.
The optical length is $\sim\!52$\,ns on average.

\begin{figure}[!t]
\centering
\includegraphics[width=\columnwidth]{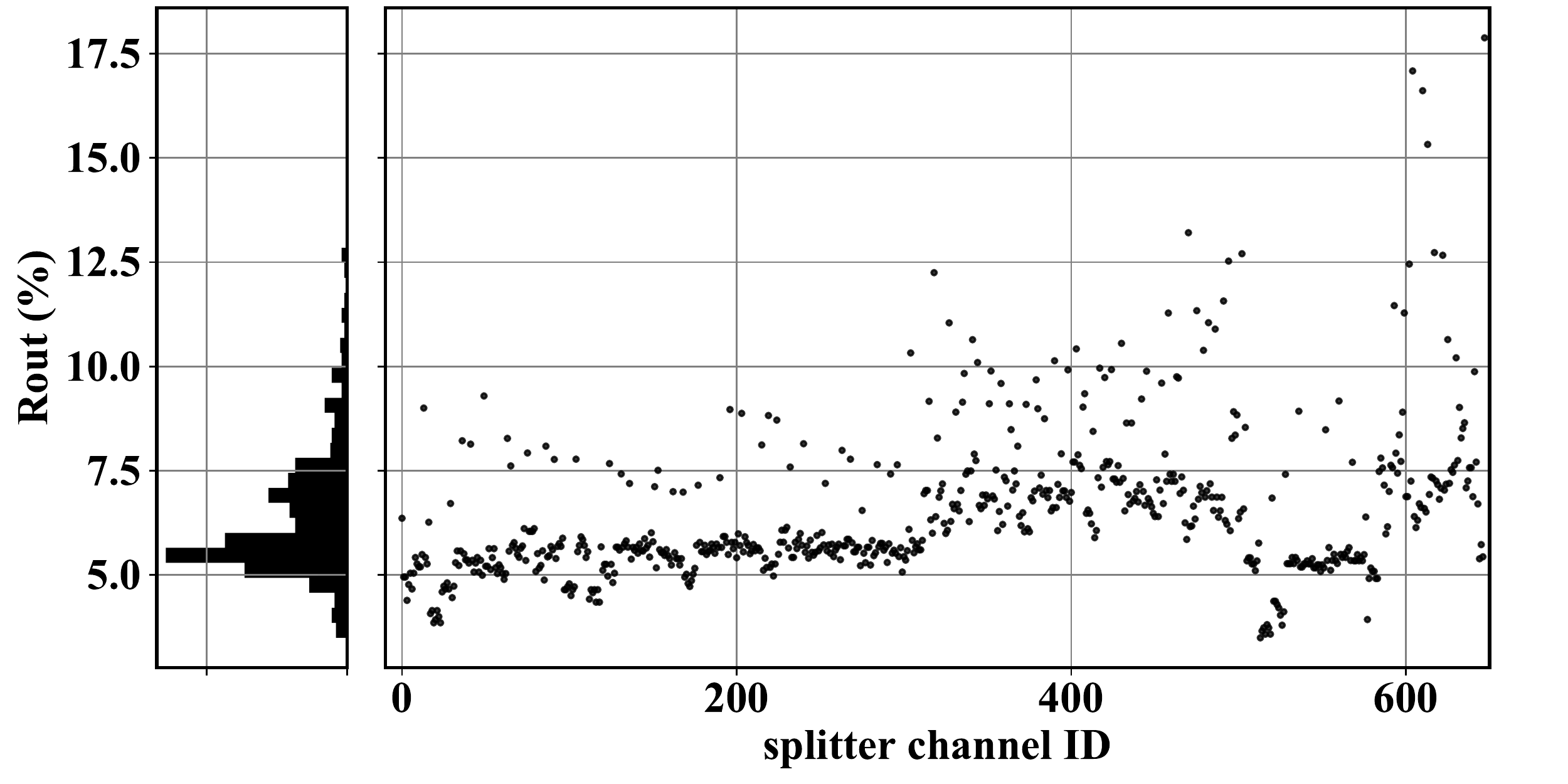}
\caption{Distribution of $R_\mathrm{out}$ of the $1\times 8$ splitter channels.}
\label{splitter_power_py}
\end{figure}
\begin{figure}[!t]
\centering
\includegraphics[width=\columnwidth]{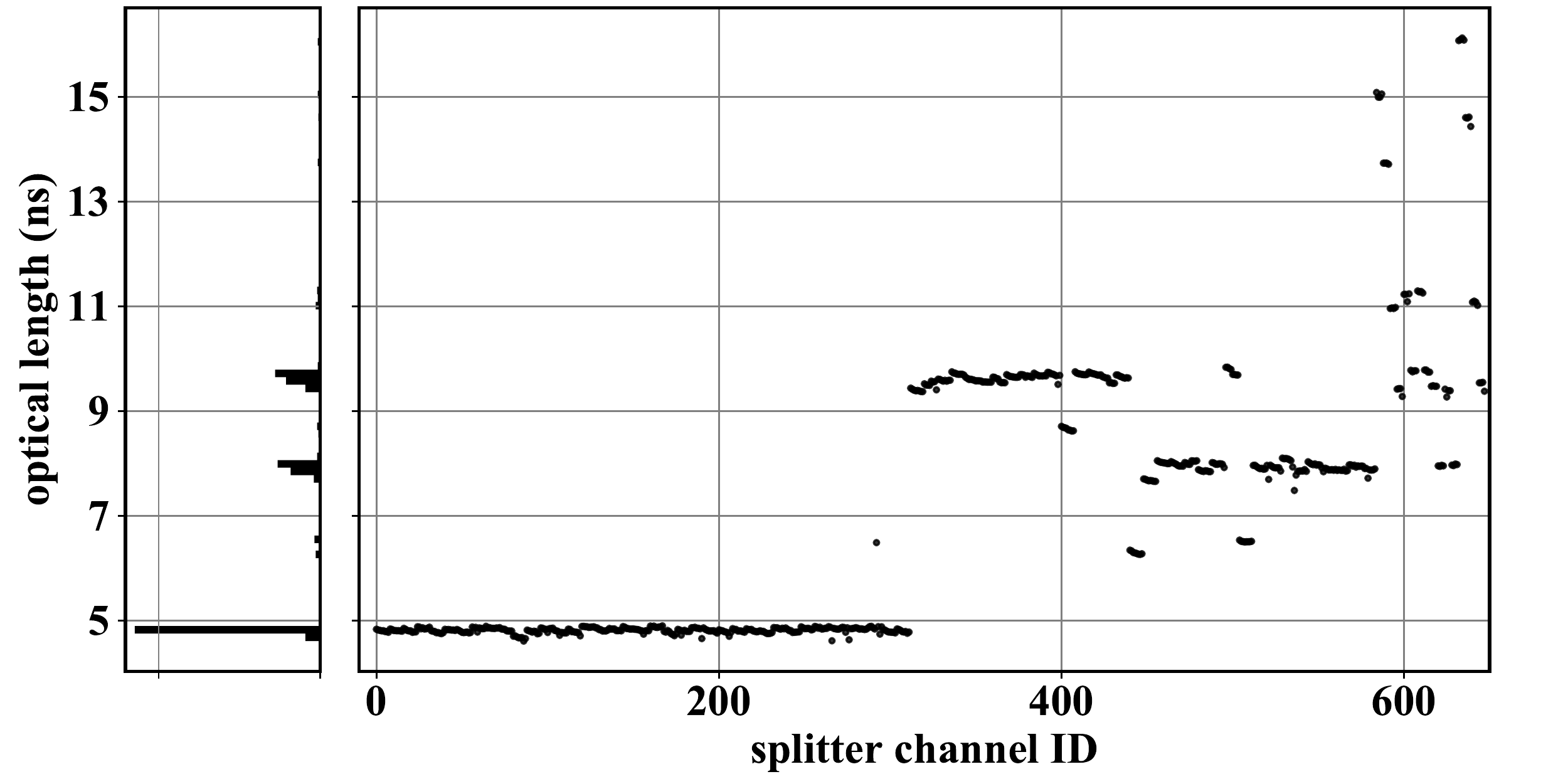}
\caption{Distribution of the optical lengths of the $1\times 8$ splitter channels.}
\label{splitter_length_py}
\end{figure}

\paragraph{$\mathit{1\times 8}$ splitter}
\frefs{splitter_power_py} and \ref{splitter_length_py} show the $R_\mathrm{out}$'s and the optical lengths, respectively,  for all the 648 channels of the 81 $1\times 8$ splitters.
The mean $R_\mathrm{out}$ is $\sim\!6$\%.
We rejected the channels below $R_\mathrm{out}\sim 4$\% from the final use to get enough power.

In each splitter, one of the outputs has a significantly larger output than the others,
as observed in \cite{Bertoni2016}. This is due to 
the primary fibre in the fusion processing having a larger output than the others. 
To make the splitting ratio more even increases the excess loss.

The channels whose IDs are in the range of $\sim$320--500 have relatively larger outputs than the others. 
We also see three groups with different optical lengths in \fref{splitter_length_py}.
These groups correspond to three different production lots.
The channels whose IDs are larger than $\sim$580 are spare splitters which were purchased $\sim$2 years later compared with other splitters.
A part of them has different optical length and power due to the different production lot.

One remarkable result we obtained is a power enhancement in the staged combination:
when we connect the first stage and the second stage splitters, one out of the $8\times8$ channels have a larger output than expected from the product of the two individual $R_\mathrm{out}$'s.
The reason for the enhancement is not clear. 
The enhanced channels are not used to get the better uniformity of outputs.

\begin{figure}[!t]
\centering
\includegraphics[width=\columnwidth]{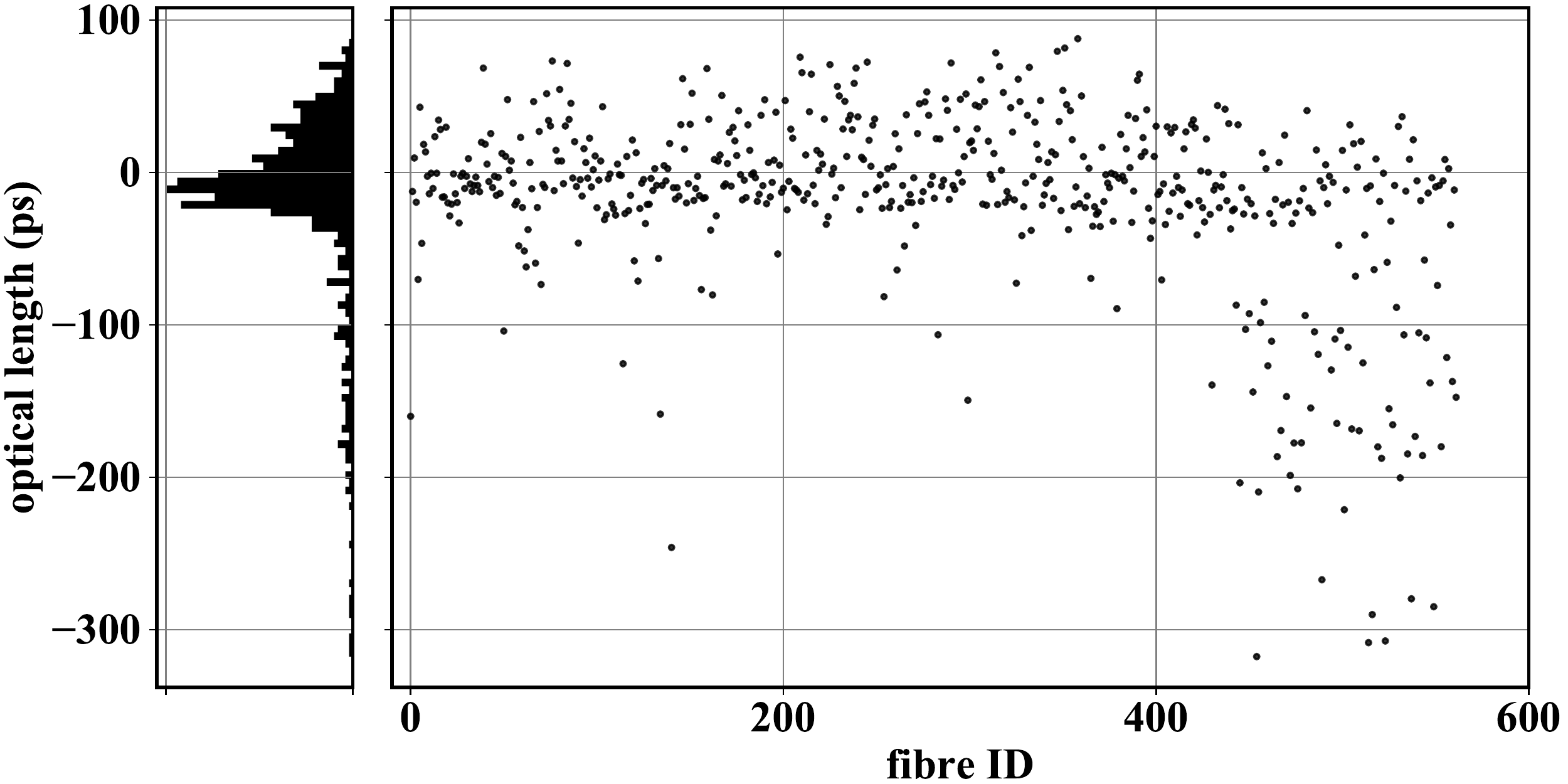}
\caption{Distribution of the optical lengths of the short fibres.}
\label{short_fibre_length_py}
\end{figure}

\paragraph{Short fibres}
The optical lengths of short fibres relative to a reference fibre are shown in \fref{short_fibre_length_py}.
Most of the differences are in the 0--100\,ps range. The fibres whose IDs are larger than 400 have
larger (negative) differences due to a difference of the production lots.

\begin{figure}[!t]
\centering
\includegraphics[width=\columnwidth]{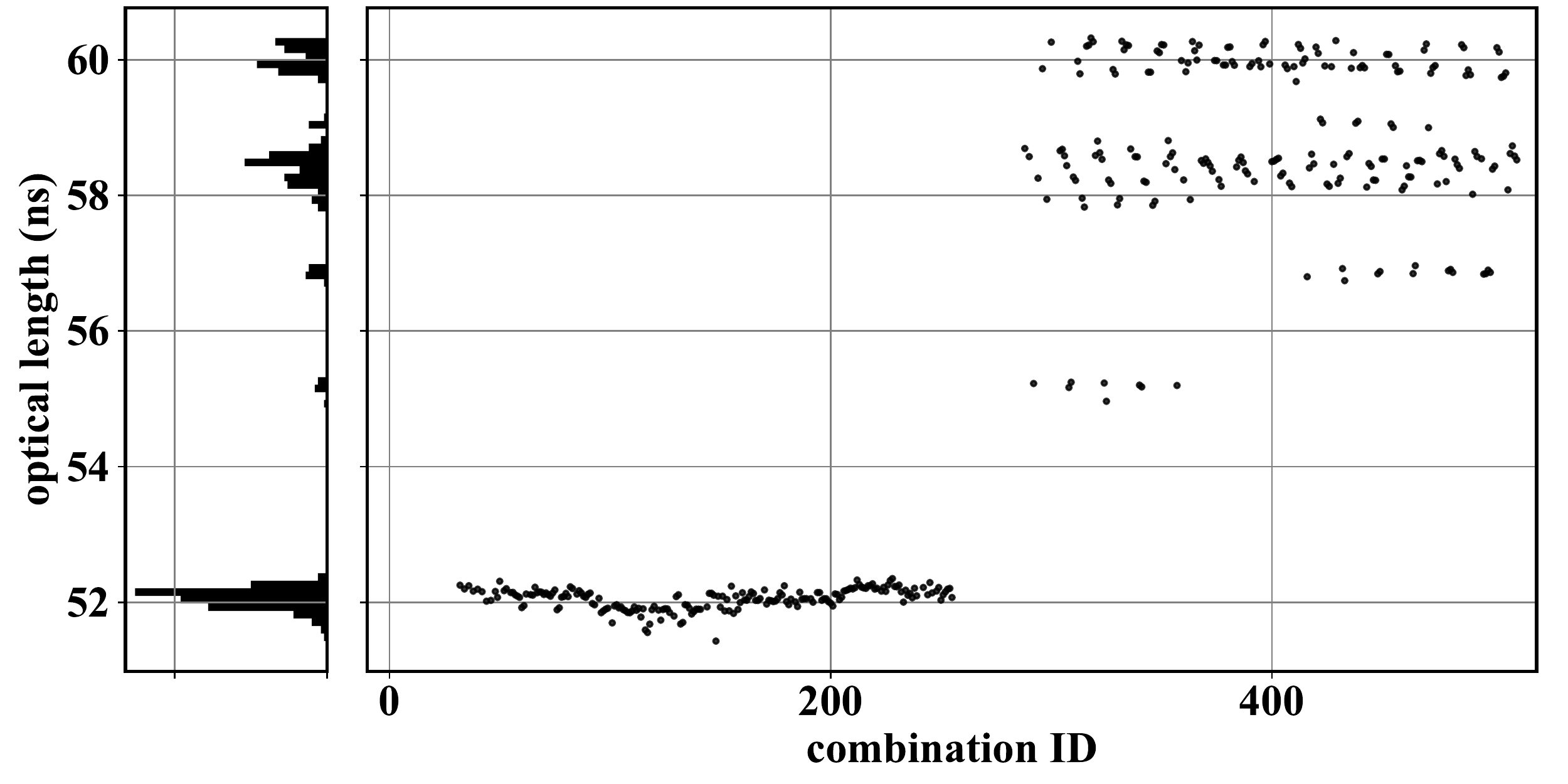}
\caption{Distribution of the overall optical lengths including $1\times8$ splitter, 2.5 m fibre, and counter.}
\label{FCLT_py}
\end{figure}

\paragraph{Overall optical length}
\label{sec:FCLT}
The overall optical lengths are measured with the following final combination: the first and the second splitters, the short fibre, and the counter.
The results are shown in \fref{FCLT_py}.
The lengths are grouped around a few different values and this is mainly due to the optical lengths of the splitters.
These values are used to calculate the time offsets in \sref{sec:determine_cal_const}.

\section{Commissioning and performance}
\label{sec:pER2017}
\subsection{Commissioning in 2017}
\label{sec:pER2017setup}
\begin{figure}[!t]
\centering
\includegraphics[width=\columnwidth]{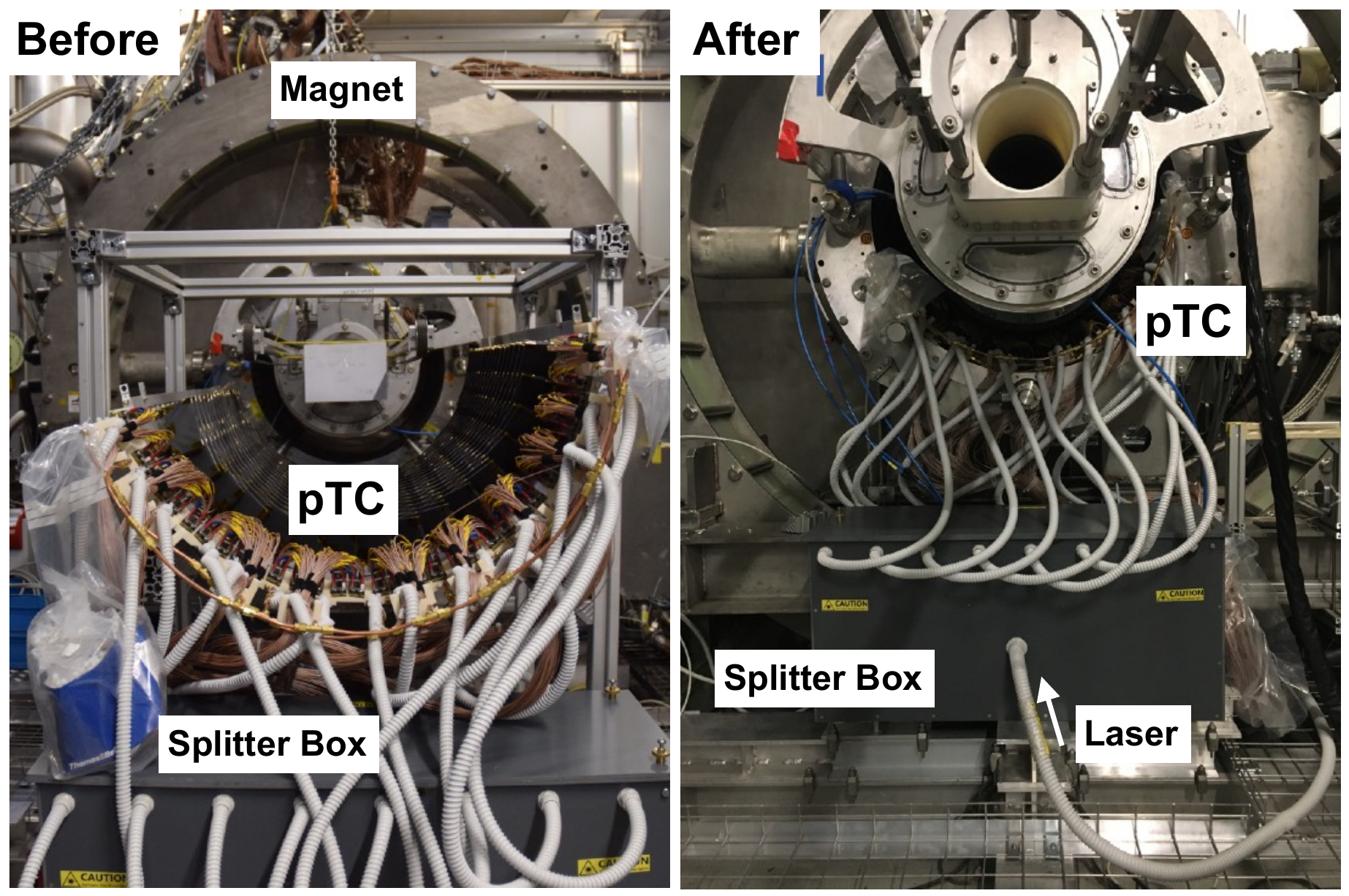}
\caption{Before-and-after pictures of installation of the downstream pTC and the end part of the laser calibration system. Optical fibres are put inside the white tubes. The detector is slid into the magnet (pointing into the paper) together with the splitter box. The pTC is not visible after installation.}
\label{Installation}
\end{figure}

In 2017\footnote{In 2016 we installed one-fourth of the detector and a part of the calibration system to 
check its basic functionality~\cite{Nakao2018}.}, we assembled the whole pTC detector, 
including the laser calibration system,  and installed it into the $\pi$E5 beam-line of the PSI proton 
accelerator complex, where the MEG II experiment will be located, to perform a commissioning run (see \fref{Installation})\footnote{Performance evaluation of the pTC can be found in \cite{Cattaneo2019}}.
The DAQ system described in \sref{sec:daq} was also installed, commissioned, and used to take pTC data.
We operated the system for two months and collected data with Michel positrons ($\mu^+\to e^+\nu_e\bar{\nu}_{\mu}$) as well as laser data.

In this commissioning run, the optical switch was not used; instead, we manually switched the eight subsets.
Laser runs with one of the eight subsets were regularly taken (at least once per day) over the period.
In addition, the laser data were taken during the Michel positron run with a mixed trigger configuration.
The full set of laser runs were taken more sparsely, roughly once per a few days, depending on other activities.   

The laser signal was not detected on a small number (2 in upstream and 9 in the downstream sector) of counters. The short fibres turned out to be broken during the installation work.  
After the commissioning run, we replaced the broken fibres with spares and the time offsets were corrected accordingly towards the next run.
This experience shows that the proposed method copes with replacement work in the future.
The installation procedure was reviewed to protect the fibres.

The laser system was used also to determine the bias voltage of the SiPMs.
We scanned the bias voltage, counter by counter, to minimise the time resolution for a fixed power laser signal.
The laser system enabled us to dynamically determine the bias voltage depending on the experimental situation, such as the detector operating temperature and the radiation damage on the SiPMs, in situ.

\subsection{Determination of the time offsets}
\label{sec:determine_cal_const}
The time offset of a counter is calculated from the mean value of $(t_1+t_2)/2 - t_{\mathrm{SYNC}}$ of 3000 laser events in a run minus the overall optical length measured in the mass test (\sref{sec:FCLT}).
The statistical uncertainty is 1\,ps.
In addition, we applied two corrections to the time offset: SiPM bias voltage correction and time-walk correction.

\subsubsection{Additional corrections}
\label{sec:additional corrections}
\begin{figure}[!t]
\centering
\includegraphics[width=\columnwidth]{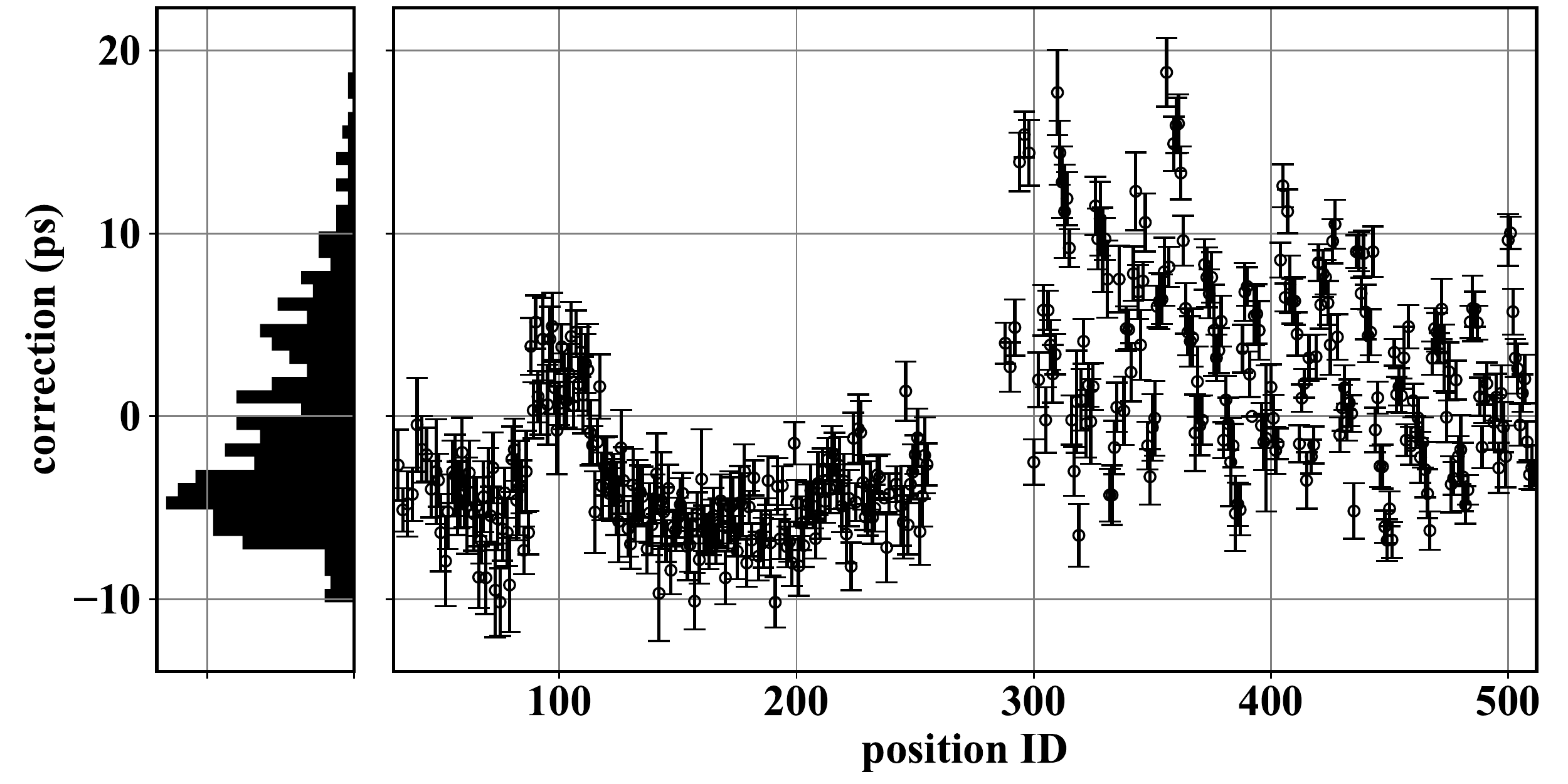}
\caption{Voltage correction of the time offsets.}
\label{voltage_correction_py}
\end{figure}

\paragraph{Voltage correction}
Since we adjust the SiPM bias voltages depending on the environmental situation as described in \sref{sec:pER2017setup}, the applied voltage can be different from the mass test. 
This affects the time offset.
Thus, we measured the difference in the time centre of laser signal at the two bias voltages and corrected the time offsets for the difference.
\fref{voltage_correction_py} shows correction values of each counter and they are distributed from $-10$\,ps to 20\,ps depending on its position.
The first half corresponds to the downstream sector and the other half corresponds to the upstream one.
The difference comes from the different radiation level because the SiPMs in the upstream were already suffered from the radiation damage in previous beam tests.
The counters whose position IDs are around 100 have different tendency compared with the others.
It is because the type of SiPMs is different from that of the others as described in \sref{sec:pTC}.
The uncertainty from this correction is estimated to be 1.5\,ps.

\paragraph{Time-walk correction}
We use a constant fraction method for the timing pickoff algorithm from the SiPM output signal to eliminate the time-walk effect.
Nevertheless, the small dependence of the time centre on the signal amplitude remains due to non-linear effects of SiPMs and electronics.
The signal amplitude of the laser data and that of the positron data can be different counter by counter, though we adjusted the laser power so that the signal amplitude is the same level as that of the positron data on average.
To evaluate the amplitude dependence of the time centre, we took amplitude scanning data by using an optical attenuator.\footnote{Since the change of the laser power changes the output light pulse shape, we used the passive attenuator (OZ OPTICS, BB-100-11-400-50/125-M-35-3S3S-3-0.5).}
The time-walk effect (time centre vs.\ signal amplitude) is measured to be $-5.74\pm0.13$\,ps/100\,mV where signal amplitude is from 200\,mV to 600\,mV.
We corrected the time offsets for the time difference between the positron data and at the laser data using its difference of the amplitude and the measured coefficient.
The uncertainty from this correction is estimated to be 4.2\,ps.

\subsubsection{Estimation of the uncertainties}
\label{sec:uncertainty}
The uncertainty related to the laser calibration is due to several components, summarised in \tref{tab:uncertainty}.
Parts of them are already discussed.
The overall uncertainty is estimated to be 27\,ps.
In this section, we focus on the two other contributions: waveform difference and variation of transit time inside the SiPMs.

\begin{table}[tb]
  \centering
   \caption{Summary of uncertainties.}
   \label{tab:uncertainty}
    \begin{tabular}{ccc} \hline
      Item & Uncertainty(ps) & Section  \\ \hline \hline
      Reproducibility & 11 & \sref{sec:reproducibility} \\ \hline
      Measurement error & \multirow{2}{*}{5.4} & \multirow{2}{*}{\sref{sec:FCLT}} \\ 
      in the mass test & & \\ \hline
      Statistics & 1.0 & \sref{sec:pER2017} \\
      Stability & 8.8 & \sref{sec:pER2017} \\
      Voltage correction & 1.5 & \sref{sec:additional corrections} \\
      Time-walk correction & 4.2 & \sref{sec:additional corrections} \\
      Waveform difference & 4.3 & \sref{sec:waveform difference} \\ \hline 
      Variation of transit time & \multirow{2}{*}{21} & \multirow{2}{*}{\sref{sec:varSiPM}} \\ 
      inside the SiPMs & & \\ \hline \hline
      Total & 27 & - \\ 
    \end{tabular}
\end{table}

\paragraph{Waveform difference}
\label{sec:waveform difference}
The waveform difference between the scintillation light and the laser light can cause systematic error on the estimation.
As shown in \fref{waveform}, the laser light has a smaller rise time than the scintillation light.
If the difference is stable over all the channels, it does not affect the time offsets because only the relative time offsets between counters matter in the time calibration.
The standard deviation of the difference over selected $\sim$250 channels was estimated to be 6.1\,ps by using the $^{90}$Sr data.
We did not apply any correction related to the waveform difference, but we set $6.1/\sqrt{2}=4.3$\,ps\footnote{by averaging over two channels} as a systematic uncertainty of the waveform difference summarised in \tref{tab:uncertainty}.

\paragraph{Variation of transit time inside SiPMs}
\label{sec:varSiPM}
\begin{figure}[!t]
\centering
\includegraphics[width=\columnwidth]{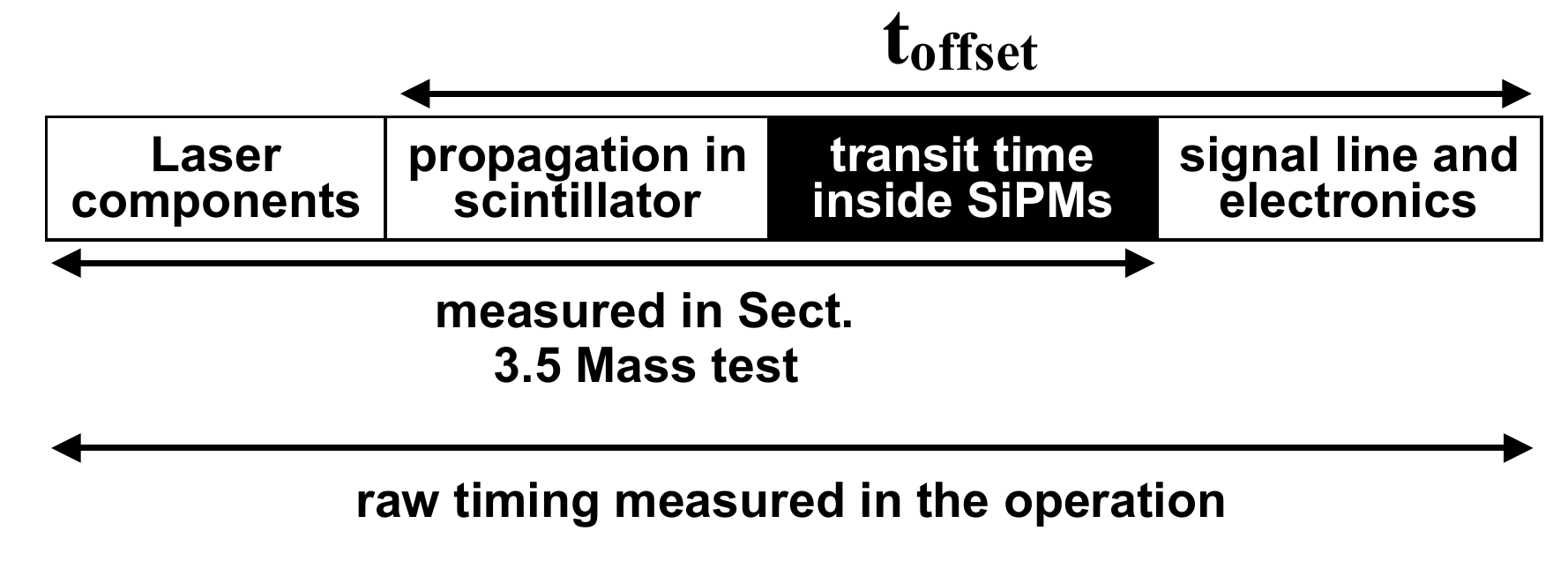}
\caption{Schematic description of the uncertainty coming from variation of transit time inside SiPMs. See \sref{sec:varSiPM} in detail.}
\label{SiPMUncert}
\end{figure}
The goal of the time calibration is to determine the time offsets $t_{\mathrm{offset}}$ written in \eref{timeoffset}.
The laser data during the operation, however, includes not only $t_{\mathrm{offset}}$ but also optical lengths of laser components shown in \fref{SiPMUncert}.
This is why we need to measure laser components beforehand in \sref{mass_test}.
If we simply subtract it from all the components in \fref{SiPMUncert}, two components are doubly subtracted: propagation in scintillator and transit time inside SiPMs.
The propagation time difference between the scintillation light and the laser light should also be included in the former
and estimated in the previous paragraph. 

The latter is estimated as follows.
We measured the hit times of electrons from a $^{90}$Sr source impinging at the centre of the counter, 
triggered by a $5\times 5\times 5~\mathrm{mm^3}$ reference counter put below the counter under test.
We repeated this measurement for $\sim$100 counters with the identical setup, and thus, the mean times of electron incidence with respect to the reference counter time are the same for all the measurements. 
The mean value of $(t_1+t_2)/2-t_{\mathrm{reference}}$, where $t_{\mathrm{reference}}$ is the timing measured on the reference counter,  shows the response of the counter, including that of the SiPMs.
The counter-by-counter variation includes the systematic errors from the setup reproducibility and the difference of scintillators, but we consider it the upper limit of the variation of transit time of SiPMs, being 21\,ps.

\subsection{Performance evaluation}

\subsubsection{Uncertainty on the time offsets}
\begin{figure}[!t]
\centering
\includegraphics[width=\columnwidth]{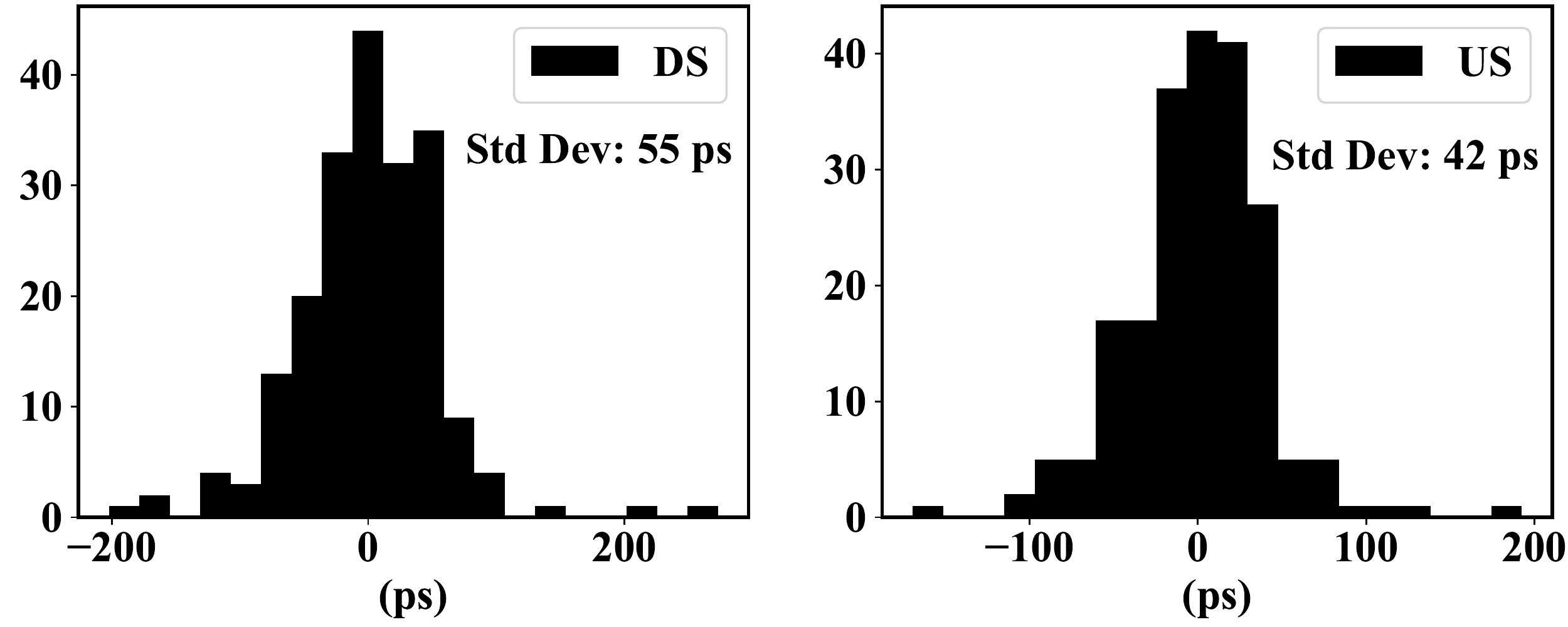}
\caption{The difference of time offsets between the laser calibration and the track-based calibration method. DS (left) and US (right) denote downstream and upstream sector, respectively.}
\label{uncertainty}
\end{figure}

In order to evaluate the uncertainty on the time offsets from the data, the results of the laser calibration were compared with those from the other time calibration method that uses positron tracks.
\fref{uncertainty} shows the difference of time offsets obtained with the two methods; they are in good agreement with a standard deviation of 48\,ps averaged over the downstream and upstream counters.

This value includes the intrinsic uncertainty of the laser calibration estimated in \sref{sec:uncertainty}, and that of the track-based calibration.
It is hard to separate these effects. In the end, these two methods are to be combined to finally determine the time offsets.
The larger value compared with the intrinsic uncertainty of the laser calibration implies a possibility of improvement of the time calibration methods.
As a conservative estimation, we conclude that the uncertainty of the determination of the time offsets is less than 48\,ps.

The effect of the time calibration on the time resolution of the pTC is estimated conservatively as follows:
The time resolution of the pTC is 38\,ps for positrons with nine hits on scintillator counters.
The uncertainty of the time calibration is randomly distributed among the hit counters.
Thus, it is smeared with the number of hit counters by averaging over all the hit counters.
Therefore, the time resolution including the effect of time calibration is estimated to be 41\,ps:
\bea
38\,\mathrm{ps}\to38\,\mathrm{ps} \oplus \frac{48\,\mathrm{ps}}{\sqrt{9}} =  41\,\mathrm{ps}.
\eea

\subsubsection{Stability and timing monitoring}
To evaluate the stability and the monitoring precision, we traced the evolution of the relative time offsets during a one-month period when the electronics and detector operation conditions were stabilised.
The mean RMS spread for the one-month period for $\sim$60 monitored counters is 8.8\,ps.
The small histogram in \fref{stability} shows its distribution and one typical example is picked up (white arrow) and shown below, in which the average of time offset  over time is set to zero.

This value includes the stability of the detector and that of the laser system.
Thus, we conclude that the stability of the time offset calculated from the laser calibration system is 8.8\,ps and the stability of the laser calibration system itself is less than 8.8\,ps.
For the timing monitoring purpose, we can detect anomalies more than $\sim$10\,ps by using the laser calibration system. 
\begin{figure}[!t]
\centering
\includegraphics[width=\columnwidth]{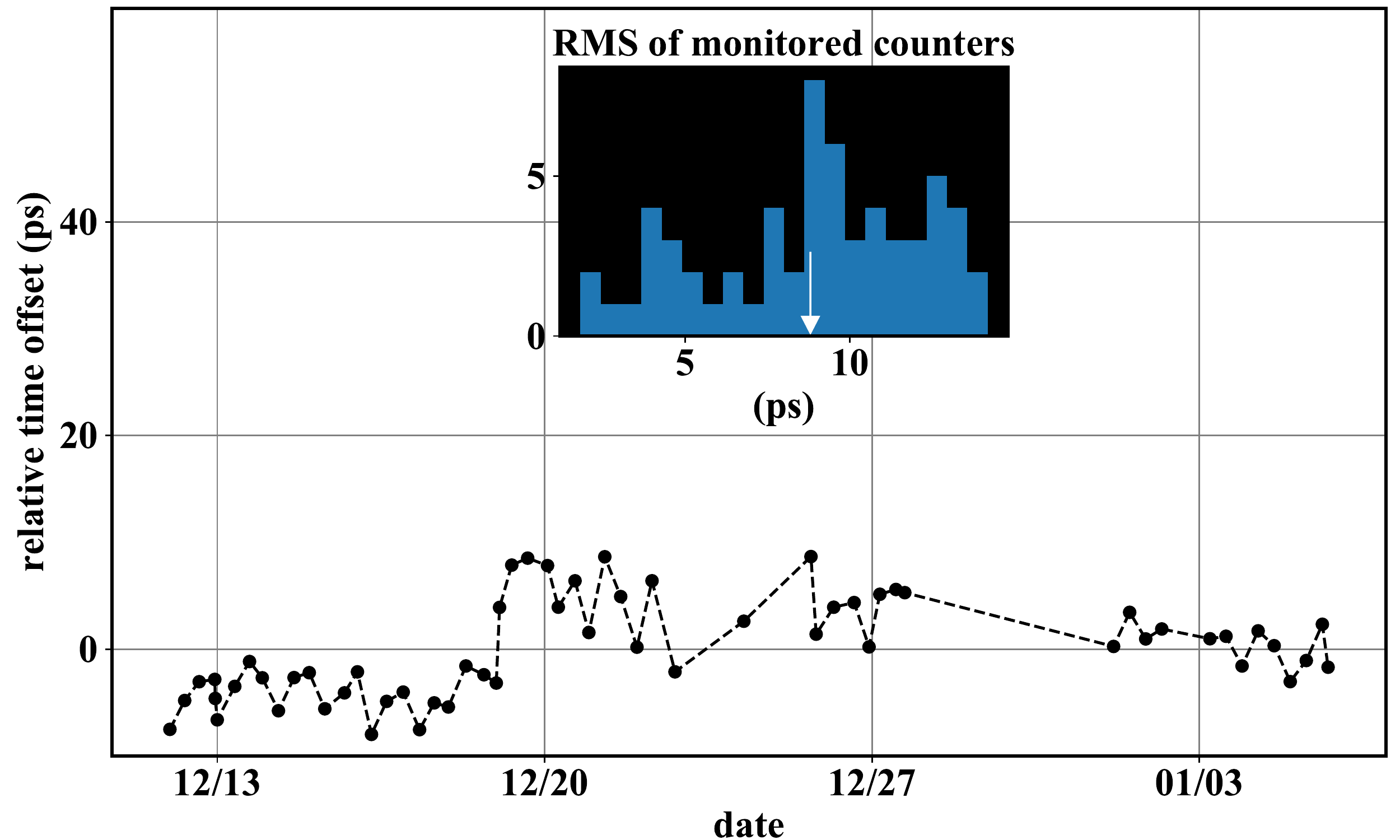}
\caption{The history of time offsets for the one-month data taking period. One typical counter is picked up from the upper histogram. The RMS of this counter is 8.8\,ps.}
\label{stability}
\end{figure}

\subsection{Prospects}
Since the optical switch was not used in 2017, we installed and tested it in 2018.
The functionalities, such as switching, the remote control, the reproducibility, and the insertion losses, were validated.
However, we found gradual degradation of output power over time and we send it back to the company for a detailed investigation.
It will be fixed or replaced towards the coming operation.

One possible upgrade on the current laser system is to introduce an asymmetric 1$\times$2 splitter instead of the current one with an equal splitting ratio.
By using one with a splitting ratio 80:20, for example, we can deliver the larger laser power to the counters while delivering sufficient power to the photo-diode for the monitoring purpose.

We reported the time resolution of the laser signal being $\sim$50\,ps.
It is good enough for our usage, but there is a way to improve it.
An MC study shows that the time resolution was dominated by how to deliver the laser light to the SiPMs inside the scintillator and the direct light from the insertion point is crucial.
Therefore different fibre insertion methods which can more directly deliver laser light to the SiPMs with less reflection can improve the time resolution.

Another possible usage of the laser calibration system is monitoring of the performance of SiPMs.
Radiation damage on SiPMs increases dark currents and worsens the time resolution\cite{Usami2019}.
We can measure the time resolution of the counters at any time in situ without having a beam time using the laser calibration system and 
estimate the effect of the radiation. 
In addition, we can also monitor the gain of the SiPMs using a fixed laser power.

\section{Conclusion}
\label{sec:conclusion}
A laser-based time calibration system for the MEG II pTC has been proposed and demonstrated.
All the optical components in the system are commercially available and easily maintained.
We have successfully developed and commissioned the system in the experimental environment.
The system shows good stability with drift smaller than 8.8\,ps and we can monitor and detect 
anomalies related to the detector and DAQ system larger than $\sim$10\,ps.
The uncertainty on the determination of time offsets is estimated to be 48\,ps and it has a moderate effect 
on the pTC time resolution.

The proposed system provides a precise timing alignment method for SiPM-based timing detectors.

\section*{Acknowledgement}
We are grateful for the support by the technical and engineering staff at PSI and our institutes, and summer students from Kyushu University and the Graduate University for Advanced Studies.
This work was supported by JSPS KAKENHI Grant Numbers JP26000004, JSPS17J04114, and JSPS Core-to-Core Program, A. Advanced Research Networks.

\bibliography{refs}

\begin{thebibliography}{10}
\expandafter\ifx\csname url\endcsname\relax
  \def\url#1{\texttt{#1}}\fi
\expandafter\ifx\csname urlprefix\endcsname\relax\def\urlprefix{URL }\fi
\expandafter\ifx\csname href\endcsname\relax
  \def\href#1#2{#2} \def\path#1{#1}\fi

\bibitem{Uchiyama2017}
Y.~Uchiyama {\it et~al}., {30-ps time resolution with segmented scintillation
  counter for MEG II}, Nucl. Instruments Methods Phys. Res. Sect. A Accel.
  Spectrometers, Detect. Assoc. Equip. 845 (2017) 507--510, \href
  {https://doi.org/10.1016/j.nima.2016.06.072}
  {\path{doi:10.1016/j.nima.2016.06.072}}.

\bibitem{MEG2}
A.~M. Baldini {\it et~al}., {The design of the MEG II experiment}, Eur. Phys.
  J. C 78~(5) (2018) 380, \href
  {https://doi.org/10.1140/epjc/s10052-018-5845-6}
  {\path{doi:10.1140/epjc/s10052-018-5845-6}}.

\bibitem{Brown1984}
J.~Brown {\it et~al}.,
  \href{https://linkinghub.elsevier.com/retrieve/pii/0167508784900589}{{The
  mark III time-of-flight system}}, Nucl. Instruments Methods Phys. Res.
  221~(3) (1984) 503--522, \href {https://doi.org/10.1016/0167-5087(84)90058-9}
  {\path{doi:10.1016/0167-5087(84)90058-9}}.

\bibitem{Kishida1987}
T.~Kishida {\it et~al}.,
  \href{https://www.sciencedirect.com/science/article/pii/0168900287906863}{{A
  laser calibration system for the KEK TOPAZ barrel TOF counters. Its
  performance and the characteristics of its major components}}, Nucl.
  Instruments Methods Phys. Res. Sect. A Accel. Spectrometers, Detect. Assoc.
  Equip. 254~(2) (1987) 367--372, \href
  {https://doi.org/10.1016/0168-9002(87)90686-3}
  {\path{doi:10.1016/0168-9002(87)90686-3}}.

\bibitem{Benlloch1990}
J.~M. Benlloch {\it et~al}.,
  \href{https://www.sciencedirect.com/science/article/pii/016890029090548K}{{Physical
  properties of the TOF (Time of Flight) scintillation counters of Delphi}},
  Nucl. Instruments Methods Phys. Res. Sect. A Accel. Spectrometers, Detect.
  Assoc. Equip. 290~(2-3) (1990) 327--334, \href
  {https://doi.org/10.1016/0168-9002(90)90548-K}
  {\path{doi:10.1016/0168-9002(90)90548-K}}.

\bibitem{Lacasse1998}
R.~Lacasse {\it et~al}.,
  \href{https://www.sciencedirect.com/science/article/pii/S0168900298001703}{{A
  time-of-flight hodoscope for the E877 spectrometer}}, Nucl. Instruments
  Methods Phys. Res. Sect. A Accel. Spectrometers, Detect. Assoc. Equip.
  408~(2-3) (1998) 408--424, \href
  {https://doi.org/10.1016/S0168-9002(98)00170-3}
  {\path{doi:10.1016/S0168-9002(98)00170-3}}.

\bibitem{Bonesini2003}
M.~Bonesini {\it et~al}.,
  \href{http://ieeexplore.ieee.org/document/1221921/}{{Laser-based calibration
  for the HARP time of flight system}}, IEEE Trans. Nucl. Sci. 50~(4 II) (2003)
  1053--1058, \href {https://doi.org/10.1109/TNS.2003.814545}
  {\path{doi:10.1109/TNS.2003.814545}}.

\bibitem{Harris2008}
F.~A. Harris {\it et~al}.,
  \href{https://www.sciencedirect.com/science/article/pii/S0168900208007109}{{BES3
  time-of-flight monitoring system}}, Nucl. Instruments Methods Phys. Res.
  Sect. A Accel. Spectrometers, Detect. Assoc. Equip. 593~(3) (2008) 255--262,
  \href {https://doi.org/10.1016/j.nima.2008.05.009}
  {\path{doi:10.1016/j.nima.2008.05.009}}.

\bibitem{Staric2017}
M.~Stari{\v{c}},
  \href{https://www.sciencedirect.com/science/article/pii/S0168900217305016?via{\%}3Dihub}{{Alignment
  and calibration methods for the Belle II TOP counter}}, Nucl. Instruments
  Methods Phys. Res. Sect. A Accel. Spectrometers, Detect. Assoc. Equip. 876
  (2017) 260--264, \href {https://doi.org/10.1016/j.nima.2017.04.038}
  {\path{doi:10.1016/j.nima.2017.04.038}}.

\bibitem{Bertoni2016}
R.~Bertoni {\it et~al}.,
  \href{http://stacks.iop.org/1748-0221/11/i=05/a=P05024?key=crossref.c8517ccf88475bf9137c1cc8e159153b}{{A
  laser diode based system for calibration of fast time-of-flight detectors}},
  J. Instrum. 11~(05) (2016) P05024--P05024, \href
  {https://doi.org/10.1088/1748-0221/11/05/p05024}
  {\path{doi:10.1088/1748-0221/11/05/p05024}}.

\bibitem{SaintGobain}
{Saint-Gobain Ceramics {\&} Plastics}, Inc.,
  \href{www.crystals.saint-gobain.com}{{CRYSTALS General Technical Data-Base
  Polyvinyltoluene}}.
\newblock
\newblock available at \special {html:<a href="www.crystals.saint-gobain.com">
  }\url{www.crystals.saint-gobain.com}\special {html:</a>}(accessed on
  2019-05-03) .

\bibitem{Lerche}
R.~Lerche, D.~Phillon, \href{http://ieeexplore.ieee.org/document/258899/}{{Rise
  time of BC-422 plastic scintillator $< 20$ ps}}, in: Conf. Rec. 1991 IEEE
  Nucl. Sci. Symp. Med. Imaging Conf., IEEE, pp. 167--170, \href
  {https://doi.org/10.1109/NSSMIC.1991.258899}
  {\path{doi:10.1109/NSSMIC.1991.258899}}.

\bibitem{ESR}
3M,
  \href{https://multimedia.3m.com/mws/media/1245089O/3m-enhanced-specular-reflector-films-3m-esr-tech-data-sheet.pdf}{{Enhanced
  Specular Reflector Films}}.
\newblock
\newblock available at \special {html:<a
  href="https://multimedia.3m.com/mws/media/1245089O/3m-enhanced-specular-reflector-films-3m-esr-tech-data-sheet.pdf">
  }\url{https://multimedia.3m.com/mws/media/1245089O/3m-enhanced-specular-reflector-films-3m-esr-tech-data-sheet.pdf}\special
  {html:</a>}(accessed on 2019-05-03) .

\bibitem{DuPont}
DuPont,
  \href{https://www.dupont.com/products-and-services/membranes-films/pvf-films.html}{{Polyvinyl
  Fluoride Films (PVF)}}.
\newblock
\newblock available at \special {html:<a
  href="https://www.dupont.com/products-and-services/membranes-films/pvf-films.html">
  }\url{https://www.dupont.com/products-and-services/membranes-films/pvf-films.html}\special
  {html:</a>}(accessed on 2019-08-08) .

\bibitem{Nakao2019}
M.~Nakao {\it et~al}., \href{http://arxiv.org/abs/1905.06270}{{A Laser-based
  Time Calibration System for the MEG II Timing Counter}}, 2019, \href
  {http://arxiv.org/abs/1905.06270} {\path{arXiv:1905.06270}}.

\bibitem{laser}
{HAMAMATSU PHOTONICS K.K.},
  \href{https://www.hamamatsu.com/resources/pdf/sys/SOCS0003E\_PLP-10.pdf}{{PLP-10
  Laser diode head Series FEATURES}}.
\newblock
\newblock available at \special {html:<a
  href="https://www.hamamatsu.com/resources/pdf/sys/SOCS0003E\_PLP-10.pdf">
  }\url{https://www.hamamatsu.com/resources/pdf/sys/SOCS0003E\_PLP-10.pdf}\special
  {html:</a>}(accessed on 2019-05-03) .

\bibitem{fibre25}
{OZ OPTICS},
  \href{https://shop.ozoptics.com/multi-mode-patchcords}{{QMMJ-31-IRVIS-50/125-1HYWT-2.5-SP}}.
\newblock
\newblock available at \special {html:<a
  href="https://shop.ozoptics.com/multi-mode-patchcords">
  }\url{https://shop.ozoptics.com/multi-mode-patchcords}\special
  {html:</a>}(accessed on 2019-05-01) .

\bibitem{fibre10}
{OZ OPTICS},
  \href{https://shop.ozoptics.com/multi-mode-patchcords}{{MMJ-33-IRVIS-50/125-3-10}}.
\newblock
\newblock available at \special {html:<a
  href="https://shop.ozoptics.com/multi-mode-patchcords">
  }\url{https://shop.ozoptics.com/multi-mode-patchcords}\special
  {html:</a>}(accessed on 2019-05-03) .

\bibitem{mode}
{Arden PHOTONICS},
  \href{https://www.ardenphotonics.com/products/modcon-mode-controller-telecom/}{{ModCon
  Mode Controller - Telecom - Arden Photonics}}.
\newblock
\newblock available at \special {html:<a
  href="https://www.ardenphotonics.com/products/modcon-mode-controller-telecom/">
  }\url{https://www.ardenphotonics.com/products/modcon-mode-controller-telecom/}\special
  {html:</a>}(accessed on 2019-05-03) .

\bibitem{2splitter}
{OZ OPTICS},
  \href{https://shop.ozoptics.com/mm-fiber-fixed-couplers-splitters}{{FUSED-12-IRVIS-50/125-50/50-3S3S3S-3-0.25}}.
\newblock
\newblock available at \special {html:<a
  href="https://shop.ozoptics.com/mm-fiber-fixed-couplers-splitters">
  }\url{https://shop.ozoptics.com/mm-fiber-fixed-couplers-splitters}\special
  {html:</a>}(accessed on 2019-05-03) .

\bibitem{8splitter}
{Lightel Technologies Inc.},
  \href{https://lightel.com/product/5/multimode-coupler}{{MMC-18-A-EVEN-1-A-30CM-R-1}}.
\newblock
\newblock available at \special {html:<a
  href="https://lightel.com/product/5/multimode-coupler">
  }\url{https://lightel.com/product/5/multimode-coupler}\special
  {html:</a>}(accessed on 2019-05-03) .

\bibitem{PD}
THORLABS,
  \href{https://www.thorlabs.com/thorproduct.cfm?partnumber=DET02AFC/M\#ad-image-0}{{DET02AFC}}.
\newblock
\newblock available at \special {html:<a
  href="https://www.thorlabs.com/thorproduct.cfm?partnumber=DET02AFC/M\#ad-image-0">
  }\url{https://www.thorlabs.com/thorproduct.cfm?partnumber=DET02AFC/M\#ad-image-0}\special
  {html:</a>}(accessed on 2019-05-03) .

\bibitem{switch}
LEONI,
  \href{https://www.leoni-fiber-optics.com/en/products-and-services/optical-components/optical-switches/}{{fibre
  Optical Switch mol 1x12-50 $\mu$m}}.
\newblock
\newblock available at \special {html:<a
  href="https://www.leoni-fiber-optics.com/en/products-and-services/optical-components/optical-switches/">
  }\url{https://www.leoni-fiber-optics.com/en/products-and-services/optical-components/optical-switches/}\special
  {html:</a>}(accessed on 2019-05-03) .

\bibitem{Kawasaki1977}
B.~S. Kawasaki, K.~O. Hill,
  \href{https://www.osapublishing.org/abstract.cfm?URI=ao-16-7-1794}{{Low-loss
  access coupler for multimode optical fiber distribution networks}}, Appl.
  Opt. 16~(7) (1977) 1794, \href {https://doi.org/10.1364/AO.16.001794}
  {\path{doi:10.1364/AO.16.001794}}.

\bibitem{Adam2013}
J.~Adam {\it et~al}.,
  \href{http://link.springer.com/10.1140/epjc/s10052-013-2365-2}{{The MEG
  detector for $\mu^+\to e^+ \gamma$ decay search}}, Eur. Phys. J. C 73~(4)
  (2013) 2365, \href {https://doi.org/10.1140/epjc/s10052-013-2365-2}
  {\path{doi:10.1140/epjc/s10052-013-2365-2}}.

\bibitem{MIDASwebsite}
\href{https://midas.triumf.ca/MidasWiki/index.php/Main\_Page}{{MIDAS website}}.
\newblock
\newblock available at \special {html:<a
  href="https://midas.triumf.ca/MidasWiki/index.php/Main\_Page">
  }\url{https://midas.triumf.ca/MidasWiki/index.php/Main\_Page}\special
  {html:</a>}(accessed on 2019-05-03) .

\bibitem{wavedream}
L.~Galli {\it et~al}., \href{http://ieeexplore.ieee.org/document/7431218/}{{A
  new generation of integrated trigger and read out system for the MEG II
  experiment}}, in: 2014 IEEE Nucl. Sci. Symp. Med. Imaging Conf., IEEE, 2014,
  pp. 1--3, \href {https://doi.org/10.1109/NSSMIC.2014.7431218}
  {\path{doi:10.1109/NSSMIC.2014.7431218}}.

\bibitem{wavedream2}
L.~Galli {\it et~al}.,
  \href{https://www.sciencedirect.com/science/article/pii/S0168900218309033?via%3Dihub}{{WaveDAQ:
  An highly integrated trigger and data acquisition system}}, Nucl. Instruments
  Methods Phys. Res. Sect. A Accel. Spectrometers, Detect. Assoc. Equip. 936
  (2019) 399--400, \href {https://doi.org/10.1016/J.NIMA.2018.07.067}
  {\path{doi:10.1016/J.NIMA.2018.07.067}}.

\bibitem{DRS}
S.~Ritt, R.~Dinapoli, U.~Hartmann,
  \href{https://www.sciencedirect.com/science/article/pii/S0168900210006091?via{\%}3Dihub}{{Application
  of the DRS chip for fast waveform digitizing}}, Nucl. Instruments Methods
  Phys. Res. Sect. A Accel. Spectrometers, Detect. Assoc. Equip. 623~(1) (2010)
  486--488, \href {https://doi.org/10.1016/J.NIMA.2010.03.045}
  {\path{doi:10.1016/J.NIMA.2010.03.045}}.

\bibitem{Cattaneo2014}
P.~W. Cattaneo {\it et~al}.,
  \href{http://ieeexplore.ieee.org/lpdocs/epic03/wrapper.htm?arnumber=6898046}{{Development
  of High Precision Timing Counter Based on Plastic Scintillator with SiPM
  Readout}}, IEEE Trans. Nucl. Sci. 61~(5) (2014) 2657--2666, \href
  {https://doi.org/10.1109/TNS.2014.2347576}
  {\path{doi:10.1109/TNS.2014.2347576}}.

\bibitem{FSJ1}
COMMSCOPE,
  \href{https://www.commscope.com/catalog/cables/pdf/part/1342/FSJ1-50A.pdf}{{FSJ1--50A}}.
\newblock
\newblock available at \special {html:<a
  href="https://www.commscope.com/catalog/cables/pdf/part/1342/FSJ1-50A.pdf">
  }\url{https://www.commscope.com/catalog/cables/pdf/part/1342/FSJ1-50A.pdf}\special
  {html:</a>}(accessed on 2019-05-03) .

\bibitem{Nakao2018}
M.~Nakao {\it et~al}., {Results from pilot run for MEG II positron timing
  counter}, in: Springer Proc. Phys., Vol. 213, 2018, pp. 237--241, \href
  {https://doi.org/10.1007/978-981-13-1316-5_44}
  {\path{doi:10.1007/978-981-13-1316-5_44}}.

\bibitem{Cattaneo2019}
P.~Cattaneo {\it et~al}.,
  \href{https://www.sciencedirect.com/science/article/pii/S0168900218311999}{{Development
  and commissioning of the 30 ps time resolution MEG II pixelated Timing
  Detector}}, Vol. 936, North-Holland, 2019, pp. 660--662, \href
  {https://doi.org/10.1016/J.NIMA.2018.09.055}
  {\path{doi:10.1016/J.NIMA.2018.09.055}}.

\bibitem{Usami2019}
M.~Usami {\it et~al}.,
  \href{https://www.sciencedirect.com/science/article/pii/S0168900218313688}{{Radiation
  damage effect on time resolution of 6 series-connected SiPMs for MEG II
  positron timing counter}}, Nucl. Instruments Methods Phys. Res. Sect. A
  Accel. Spectrometers, Detect. Assoc. Equip. 936 (2019) 572--573, \href
  {https://doi.org/10.1016/J.NIMA.2018.10.053}
  {\path{doi:10.1016/J.NIMA.2018.10.053}}.

\end{thebibliography}
\end{document}